\documentclass[letterpaper,11pt]{article}

\usepackage{amsmath,amssymb,textcomp,graphicx,array,natbib} 
\usepackage{fullpage,rotate}
\usepackage[small,bf,up]{caption}

\newcommand{\half}{\mbox{\small $\frac{1}{2}$}}

\renewcommand{\baselinestretch}{1.0}

\begin{document}
\noindent
{\LARGE \sf Computer Model Calibration Using the Ensemble
 Kalman Filter}\\[0.5cm]
 Dave Higdon, Statistical Sciences Group, Los Alamos National Laboratory\\
 Matt Pratola, Statistical Sciences Group, Los Alamos National Laboratory\\
 Jim Gattiker, Statistical Sciences Group, Los Alamos National Laboratory\\
 Earl Lawrence, Statistical Sciences Group, Los Alamos National Laboratory\\
 Charles Jackson, University of Texas Institute for Geophysics\\
 Michael Tobis, University of Texas Institute for Geophysics\\
 Salman Habib, High Energy Physics Division, Argonne National Laboratory\\
 Katrin Heitmann, High Energy Physics Division, Argonne National Laboratory\\
 Steve Price, Fluid Dynamics Group, Los Alamos National Laboratory\\

\graphicspath{{./}{Figs/}}

\noindent
{\small
The ensemble Kalman filter (EnKF) \citep{evensen2009data} has
proven effective in quantifying uncertainty in a number of challenging dynamic,
state estimation, or data assimilation, problems such as
weather forecasting and ocean modeling.  
In these problems a high-dimensional state parameter is successively
updated based on recurring physical observations, with the aid of 
a computationally demanding forward
model that propagates the state from one time step to the next.  
More recently, the EnKF
has proven effective in history matching in the petroleum engineering
community \citep{evensen2009ensemble,oliver2010recent}.  Such applications typically
involve estimating large numbers of parameters, describing an oil reservoir,
using data from production history that
accumulate over time.  Such history matching problems are especially challenging
examples of computer model calibration since they involve a large number of 
model parameters as well as a computationally demanding forward model.  
More generally, computer model calibration 
combines physical observations
with a computational model -- a computer model -- to estimate unknown parameters in
the computer model.  This paper explores how the EnKF can be
used in computer model calibration problems, comparing it to other more common
approaches, considering applications in climate and cosmology.
\\[.2cm]
 Keywords: computer experiments; model validation; data assimilation;
 uncertainty quantification; Gaussian process; parameter estimation; 
 Bayesian statistics
}

\section{Introduction}

The ensemble Kalman filter (EnKF) has
proven effective in quantifying uncertainty in a number of challenging dynamic,
state estimation, or data assimilation, problems.  Applications include
weather forecasting \citep{houtekamer2005atmospheric}, ocean modeling
\citep{evensen2003ekf}, storm tracking \citep{aksoy2009multicase},
hydrology \citep{moradkhani2005dual} and wildfire
modeling \citep{mandel2004note}, just to name a few.  
In these data assimilation problems, a high-dimensional state parameter is successively
updated based on recurring physical observations, with the aid of 
a computationally demanding forward
model that propagates the state from one time step to the next.  
The EnKF iteratively updates an ensemble of state vectors, 
using a scheme motivated by the standard Kalman filter
\citep{meinhold1983understanding,west:harr:1997}, producing an updated ensemble 
of states that is affected by both the forward model and the physical observations.
More recently, the EnKF
has proven effective in history matching in the petroleum engineering
community \citep{evensen2009ensemble,oliver2010recent}.  Such applications typically
involve estimating large numbers of parameters, describing an oil reservoir,
using data from production history that
accumulate over time.  Such history matching problems are especially challenging
examples of computer model calibration since they involve a large number of 
model parameters as well as a computationally demanding forward model.  
Unlike standard data assimilation problems, 
here focus is on estimation of a static model parameter vector, rather than an
evolving state vector.

This paper explores how the EnKF can be
used in computer model calibration problems that typically have a collection
of model parameters to be constrained using physical observations.  
We first use a simple 1-d
inverse problem to describe standard Bayesian approaches to produce a posterior
distribution for the unknown model parameter vector, as well as the resulting model
prediction.  We then go on to describe how
the EnKF can be used to address this basic problem,
with examples taken from the literature in climate and cosmology.
%
We end with conclusions summarizing
the strengths and weaknesses of using the EnKF for computer model calibration.

\subsection{A simple inverse problem}
\label{sec:simpleInverse}

\begin{figure}[ht]
  \centerline{
\fbox{
\begin{picture}(420,180)(0,0)
\put(0,0){\includegraphics[width=3.2in,angle=0] {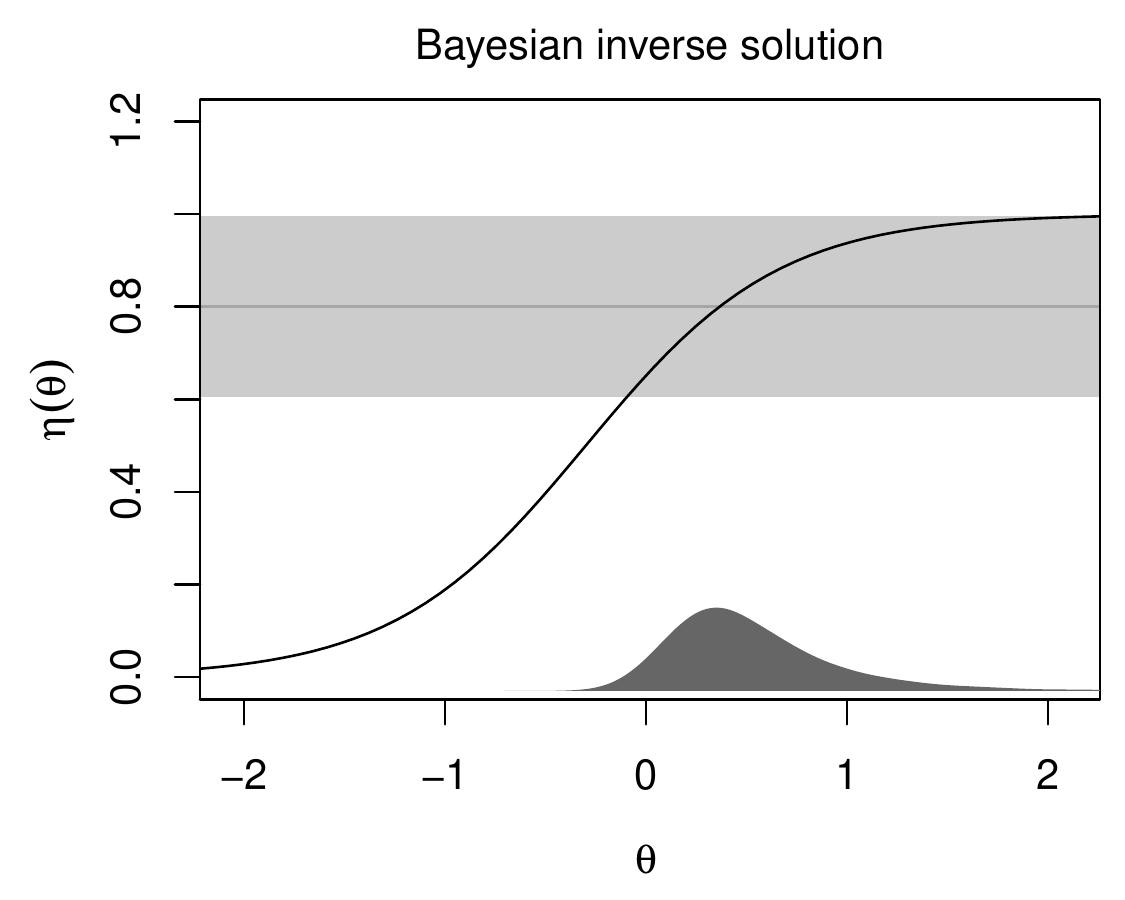}}
\put(235,90){\parbox{2.5in}{\footnotesize \sf A simple inverse problem in 
  which $\eta(\cdot)$, given by the black line, denotes the forward model, mapping the unknown 
  parameter $\theta$ into an observable $\eta(\theta)$.\\[.2cm]
  The physical observation $y$ is a noisy version of $\eta(\theta)$:
  \[
    y = \eta(\theta) + \epsilon
  \]
  where $\epsilon \sim N(0,\sigma_y^2=.1^2)$.  The horizontal gray line and band denote the
  measurement ($y=.8$) and its uncertainty $y \pm 2\sigma_y$.\\[.2cm]
  The model parameter $\theta$ is given a $N(0,1)$ prior.  The resulting posterior
  for $\theta$ is given by the shaded density at the bottom of the figure.}}
\end{picture} 
}}
   \caption{\label{fig:inverse} A simple 1-dimensional inverse problem and resulting 
   posterior density.}
\end{figure}
In order to describe the basic approaches to inverse problems, we first describe a 
simple, 1-dimensional inverse problem shown in Figure \ref{fig:inverse}.  We take
$\eta(\cdot)$ to denote the forward model.  It requires a single model parameter
$\theta$, producing a univariate output $\eta(\theta)$ that is comparable to a physical 
measurement $y$.  Here we take the sampling model for $y$ to be normally
distributed about the forward model's output when the true value of the model
parameter is input
\[
 L(y|\eta(\theta)) \propto \exp\{-\half \sigma^{-2}_y (y - \eta(\theta))^2\}
\]
where the observation error is assumed known to be $\sigma_y = .1$.

After specifying a standard normal prior for the model parameter,
with $\pi(\theta)$ denoting the prior density, the posterior
density is given by
\begin{eqnarray}
\label{eq:simple0}
 \pi(\theta|y) &\propto& L(y|\eta(\theta)) \times \pi(\theta) \\
 \nonumber
  &\propto& \exp\{-\half \sigma^{-2}_y (y - \eta(\theta))^2\} \times
     \exp\{-\half \theta^2\}.
\end{eqnarray}
Thus an evaluation of the posterior requires a run of the forward model.
While this simple, 1-dimensional density is trivial to evaluate, many inverse
problems have to deal with a large model parameter vector (dimensions ranging from
$10$ to $10^8$) as well as a computationally demanding forward model that may
take a long time to evaluate (our experience ranges from seconds to weeks). 

\subsubsection{Using a Gaussian process emulator}
While Markov chain Monte Carlo (MCMC) remains a popular approach for exploring the
resulting posterior distribution \citep{kaip:some:2004,tarantola2005inverse}, the demands required by the size of the model parameter
vector and the computational demands of the forward model have inspired recent research
focused on overcoming these hurdles.  These research efforts
range from response surface approximation
of the forward model $\eta(\cdot)$ \citep{kenn:ohag:2001,higd:kenn:cave:2005}, to constructing
reduced, or simplified forward models \citep{galbally2010non,lieberman2010parameter}, to
polynomial chaos approximations of the prior model 
\citep{ghanem2006construction,marzouk2009dimensionality}, to exploiting
multiple model fidelities \citep{christen2005mcm,efendiev2009efficient}.

\begin{figure}[ht]
  \centerline{
\fbox{
\begin{picture}(420,180)(0,0)
\put(0,0){\includegraphics[width=3.2in,angle=0] {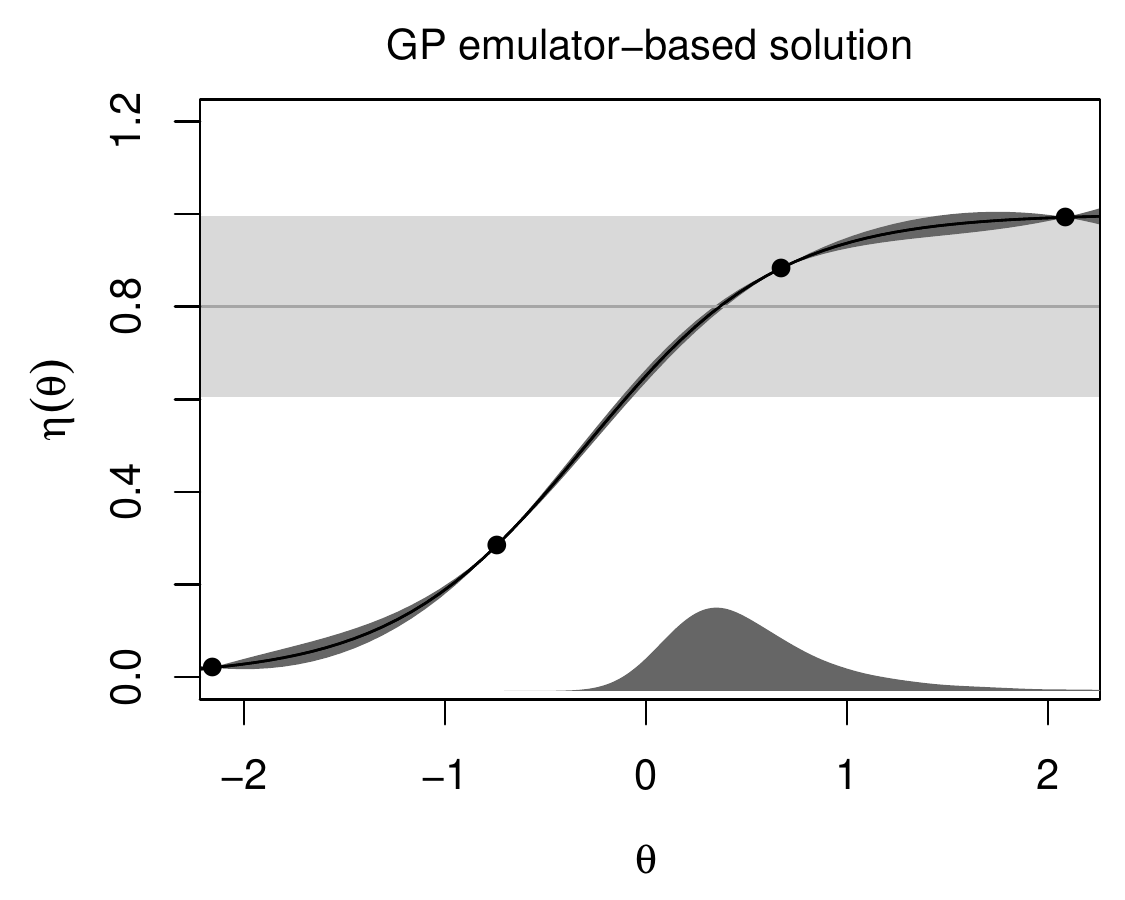}}
\put(235,90){\parbox{2.5in}{\footnotesize \sf The forward model $\eta(\cdot)$ is run at
4 values of the input parameter, producing model observations $\eta^\circ$ (black dots).  
A GP prior $\pi(\eta(\cdot))$ is specified
for the forward model.
\begin{eqnarray*}
\mbox{likelihood (measurement) } & y \sim N(\eta(\theta),\sigma_y^{2}) \\
\mbox{prior for $\eta(\cdot)$ } & \pi(\eta(\cdot)) \\
\mbox{likelihood (simulations) } & \eta^\circ \sim \pi(\eta(\cdot)) \\
\mbox{prior for $\theta$ } & \pi(\theta) \\
\Rightarrow \mbox{posterior  } & \pi(\theta,\eta(\cdot)|y,\eta^\circ)
\end{eqnarray*}
Conditioning on both the physical observation (light, horizontal band) and the four
model runs, the posterior distribution for both $\eta(\cdot)$ and $\theta$ is
produced (shaded density).
}}
\end{picture} 
}}
   \caption{\label{fig:gp} Using a Gaussian process prior for the forward model to reduce
   the number of model runs necessary for posterior exploration for the 
   simple inverse problem.}
\end{figure}

Gaussian processes (GPs) are commonly used to emulate the computer
model response, producing a probabilistic description of the response
at untried parameter settings -- see \citet{kenn:ohag:2001} or
\citet{baya:berg:paul:2007} for just a couple of examples.  This basic approach
is depicted in Figure \ref{fig:gp}.  In this case, a GP prior is used to model the unknown
function $\eta(\cdot)$ and a collection of forward model runs
$\eta^\circ=(\eta(\theta_1^\circ),\ldots,\eta(\theta_m^\circ))'$, over a 
collection of input parameter settings $\theta^\circ=(\theta_1^\circ,\ldots,\theta_m^\circ)'$, 
are used to infer the forward model response
at untried input parameter settings.

Thus  the basic fomulation (\ref{eq:simple0}) is augmented to incorporate this
GP prior for the forward model
\[
\eta(\cdot) \sim GP(m(\cdot),C(\cdot,\cdot)),
\]
where the mean function $m(\cdot)$ may be a constant \citep{sack:welc:mitc:wynn:1989,
kenn:ohag:2001}, or a more complicated regression function \citep{crai:gold:roug:2001,vernon2010galaxy}, and
the covariance function $C(\cdot,\cdot)$ is typically of product form, requiring just a single
additional parameter for each dimension of the input parameter.  For this simple problem,
we take the mean and covariance functions as fixed, leading to the posterior
form
\[
\pi(\theta|y,\eta^\circ) \propto 
 \exp\{-\half (v_\theta + \sigma^2_y)^{-1} (y - \mu_\theta)^2\} \times
     \exp\{-\half \theta^2\}.
\]
Here $\mu_\theta$ and $v_\theta$ are the mean and variance given by the
GP model after conditioning on the forward model runs $\eta^\circ$
\begin{eqnarray*}
 \mu_\theta &=& C(\theta,\theta^\circ)C(\theta^\circ,\theta^\circ)^{-1} (\eta^\circ - m(\theta^\circ)) + m(\theta)\\
 v_\theta &=&    C(\theta,\theta)- C(\theta,\theta^\circ)C(\theta^\circ,\theta^\circ)^{-1}C(\theta^\circ,\theta)  ,
\end{eqnarray*}
where, for example, $C(\theta^\circ,\theta^\circ)$ produces a $m \times m$ matrix whose 
$ij$ entry is $C(\theta^\circ_i,\theta^\circ_j)$ and $C(\theta,\theta^\circ)$ produces a $m$-vector
whose $j$th element is $C(\theta,\theta^\circ_j)$.
See \citet{higd:kenn:cave:2005} for
details regarding the posterior specification when the GP (and other) parameters are not taken
as fixed, and the input parameter is multivariate.

\subsubsection{Using the ensemble Kalman filter}
\label{sec:enkf}

Below we briefly describe two basic variants of the the EnKF for 
computer model calibration, differing in how they use the ensemble of model 
runs to approximate, and represent, 
the resulting posterior distribution.  In both cases, an ensemble of draws $\theta^\circ$ from the prior distribution of the model parameter are paired with the resulting simulation output to produce an ensemble of  $(\theta^\circ,\eta(\theta^\circ))$ pairs, from which the sample covariance is used to produce an approximation to the posterior distribution.  Hence we treat the input parameter settings $\theta^\circ_1,\ldots,\theta^\circ_m$ as $m$ draws from the prior distribution $\pi(\theta)$. Note that even though the distribution of the simulator response $\eta(\theta)$ is completely determined by the distribution for $\theta$, the EnKF uses a joint
normal model for $(\theta,\eta(\theta))$ to motivate its calculations.

Next we describe two variants of the EnKF algorithm for computer model calibration.
One uses a Gaussian representation of the posterior distribution, the other uses
an ensemble representation.

\subsubsection*{Gaussian representation}
\begin{figure}[htb]
  \centerline{
   \includegraphics[width=3.3in,angle=0] {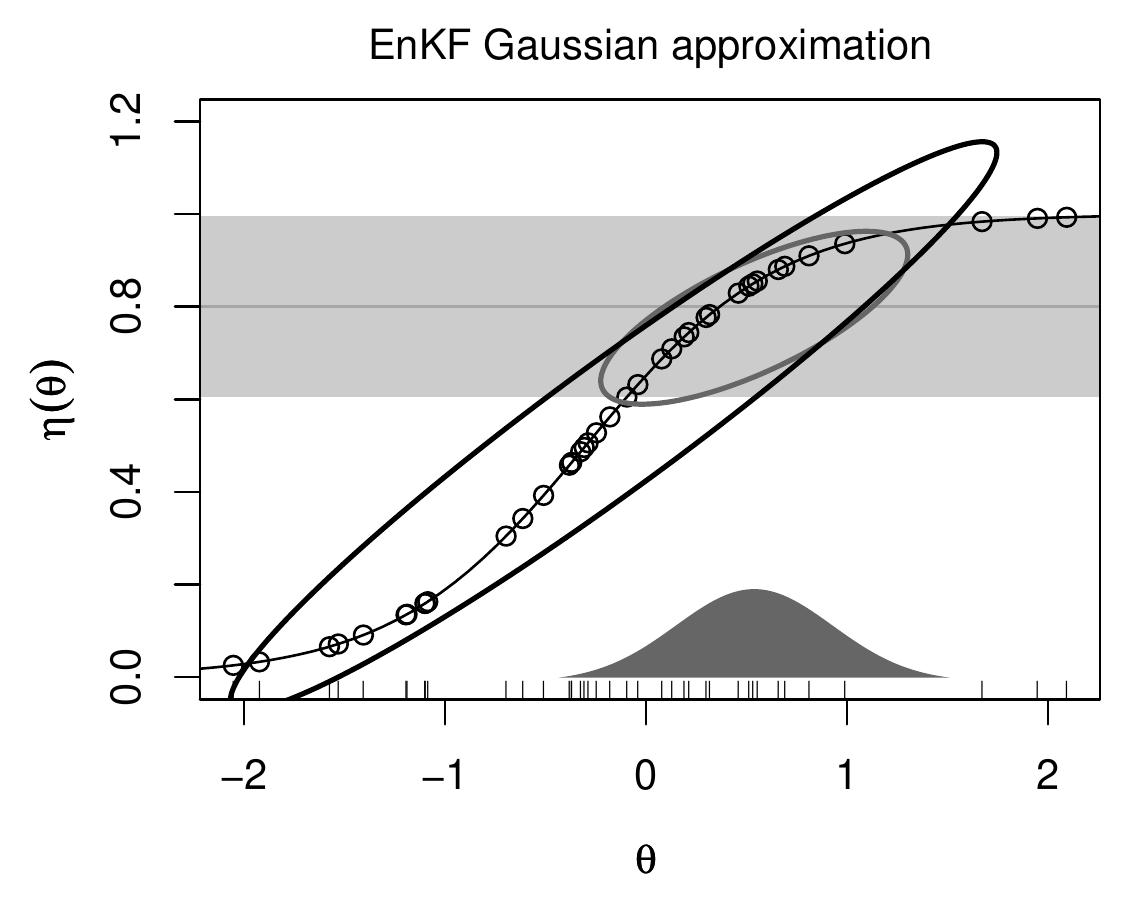}
   \includegraphics[width=3.3in,angle=0] {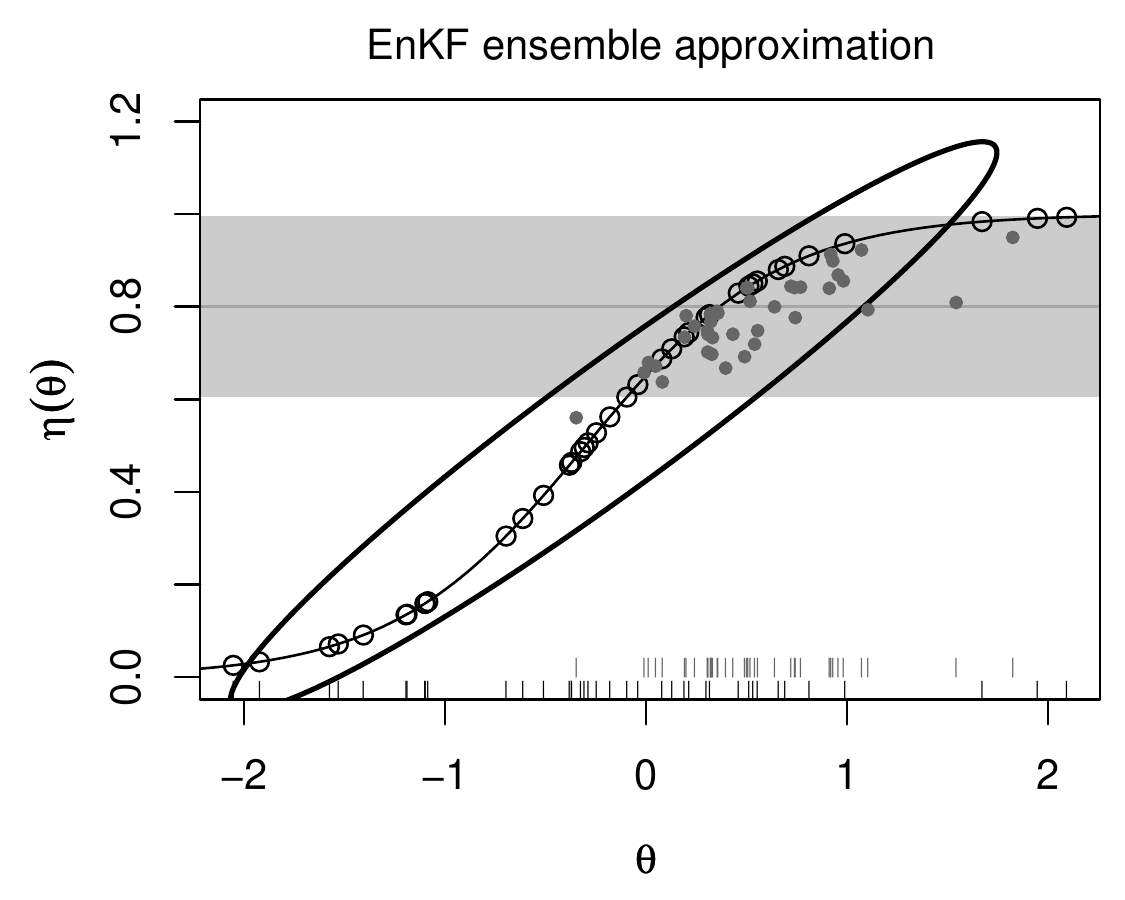}}
   \caption{\label{fig:enkf} Left: Gaussian representation of the posterior distribution for
   $(\theta,\eta(\theta))$ resulting from the ensemble
   Kalman filter (EnKF).   The approximate normal prior distribution for $(\theta,\eta(\theta))$ is 
   depicted by the black ellipse, estimated from the ensemble (circle plotting symbols).
   The resulting posterior distribution is approximated as normal, depicted by
   the gray ellipse. The marginal posterior for $\theta$ is given by the
   shaded density.
   Right: the ensemble representation of the posterior distribution for 
   $(\theta,\eta(\theta))$ resulting from the EnKF.  Here the updated sample (gray dots)
   are approximate draws from the posterior distribution.}
\end{figure}

The first approach fits a multivariate normal distribution to the 
ensemble for $(\theta^\circ,\eta(\theta^\circ))$.  
The algorithm is depicted in
the left frame of Figure \ref{fig:enkf} and described below.
\begin{enumerate}
\item
For each of the $m$ simulations form the ensemble of joint vectors
\begin{equation}
\begin{pmatrix} \theta_k^\circ \\ \eta(\theta^\circ_k) \end{pmatrix},\, k=1,\ldots,m.
\end{equation}
With these $m$ vectors, compute the sample mean vector $\mu_{\rm pr}$ and sample covariance matrix $\Sigma_{\rm pr}$.  For the simple inverse problem here,
$\mu_{\rm pr}$ is a $2$-vector and $\Sigma_{\rm pr}$ is $2 \times 2$, but this recipe
is quite general.
%
\item
In this simple inverse problem, the physical observation $y$ corresponds to the 2nd
element of the joint $(\theta,\eta(\theta))$ vector. Take ${H}$ be $(0,1)'$ to be the 
observation matrix.   The likelihood can be written
\begin{equation}
L(y|\eta(\theta)) \propto \exp\left\{ -\frac{1}{2} 
\left( y - H \begin{pmatrix} \theta \\ \eta(\theta) \end{pmatrix} \right)'
\Sigma_y^{-1}
\left( y - H \begin{pmatrix} \theta \\ \eta(\theta) \end{pmatrix} \right) \right\}.
\end{equation}
More generally the observation operator $H$ can select elements of $\eta(\theta)$ 
that are observed, or can be specified to interpolate between values of the simulator
output.
\item
Combining the normal approximation to the prior with the normal likelihood results in an updated, or posterior, distribution for $(\theta,\eta)$ for which
\begin{equation}
\label{eq:enkfgpost}
\begin{pmatrix} \theta \\ \eta \end{pmatrix}
| y
\sim
N(\mu_{\rm post},\Sigma_{\rm post}),
\end{equation}
where
\begin{equation}
\label{eq:enkfgprec}
\Sigma_{\rm post}^{-1} = \Sigma_{\rm pr}^{-1} + 
  H' \Sigma_y^{-1} H
\end{equation}
and
\begin{equation}
\label{eq:enkfgmean}
 \mu_{\rm post} = \Sigma_{\rm post} \left(
   \Sigma_{\rm pr}^{-1} \mu_{\rm pr} +
   H' \Sigma_y^{-1} y \right).
\end{equation}
Note that the posterior mean can be rewritten in a form more commonly used
in Kalman filtering
\begin{equation*}
\mu_{\rm post}
=
\mu_{\rm pr}
+ \Sigma_{\rm pr}H'
(H\Sigma_{\rm pr}H' + \Sigma_y)^{-1}
(y - H\mu_{\rm pr})
\end{equation*}
where $\Sigma_{\rm pr}H'
(H\Sigma_{\rm pr}H' + \Sigma_y)^{-1}$ is the Kalman gain
matrix.
\end{enumerate}
The joint normal computations
used here effectively assume a linear plus Gaussian noise relationship
between $\theta$ and $\eta(\theta)$, inducing a normal posterior for $\theta$.

\subsubsection*{Ensemble representation}
The second approach is basically the usual EnKF as applied to
time evolving systems, but here, only for a single time step.  The goal 
is to perturb each member of the ensemble $(\theta^\circ_k,\eta(\theta^\circ_k))$, in order to 
produce an updated member $(\eta_k^{(1)},\theta^{(1)}_k)$
whose mean and variance match the posterior produced by the Gaussian representation
EnKF described above.  This updated member is not produced with the 
simulator so that $\eta^{(1)}_k$ will not be equal to the simulator evaluated at 
updated parameter value $\eta(\theta^{(1)}_k)$.  Here we describe the 
perturbed data version of the EnKF described in \citet{evensen2009ensemble}.
A number of variants of this basic approach 
exist; see \citet{anderson2001ensemble} 
and \citet{szunyogh2008local}, for example.
The algorithm is
given below.
\begin{enumerate}

 \item Construct the sample covariance matrix $\Sigma_{\rm pr}$ as in Step 1 of the previous algorithm.

\item  For $k=1,\dots,m$ do:
   \begin{enumerate}
     \item Draw a perturbed data value $y_k \sim N(y,\Sigma_y)$.
     \item Produce the perturbed ensemble member
     \begin{equation} 
      \label{eq:enkfe1}
       \begin{pmatrix} \theta^{(1)}_k \\ \eta_k^{(1)} \end{pmatrix}
       = \Sigma_{\rm post} \left(
        \Sigma_{\rm pr}^{-1} 
          \begin{pmatrix} \theta^\circ_k \cr \eta(\theta^\circ_k) \end{pmatrix} +
       H' \Sigma_y^{-1} y_k \right).
     \end{equation}
     where $\Sigma_{\rm pr}$ and $\Sigma_{\rm post}$ are defined in the previous
     algorithm.  Note this perturbation of the ensemble member 
     can be equivalently written using the more standard Kalman gain update:
     \begin{equation}
      \label{eq:enkfe2}
       \begin{pmatrix} \theta^{(1)}_k \\ \eta_k^{(1)} 
	\end{pmatrix}
	=
	\begin{pmatrix} \theta^\circ_k \cr \eta(\theta^\circ_k)
	\end{pmatrix}
	+ \Sigma_{\rm pr}H'
	(H\Sigma_{\rm pr}H' + \Sigma_y)^{-1}
	(y_k - \eta(\theta^\circ_k))
       \end{equation}
   \end{enumerate}
   \item Treat this updated, $m$ member ensemble
   \begin{equation*}
      \begin{pmatrix} \theta^{(1)}_k \\ \eta_k^{(1)} \end{pmatrix},\, k=1,\ldots,m.
   \end{equation*}
   as draws from the updated, posterior distribution for $(\theta,\eta)$ given the
   initial ensemble $(\theta^\circ,\eta(\theta^\circ))$ and the physical observation $y$.
\end{enumerate}
This approach uses a Bayesian update of two normal forms, with each ensemble
member updated separately.  Here the normal prior is centered at the ensemble member, and the normal likelihood is centered at the perturbed data value, rather than at the ensemble mean and the actual data value.  This update sets the new ensemble
value $(\theta^{(1)}_k,\eta^{(1)}_k)$ to the mean of this resulting combination of normal
distributions.

This produces a posterior ensemble for the joint distribution of $(\theta,\eta)$, given
by the gray dots in the right hand frame of Figure \ref{fig:enkf}.  Hence the
difference between these two representations can be seen 
Figure \ref{fig:enkf} -- compare the gray ellipse in
the left frame, representing the updated normal posterior, to the gray dots in the right frame,
representing draws using this ensemble representation.  

Note that if we take $(\theta^\circ_k,\eta^\circ_k)$ to be a draw from a distribution with
mean $\mu_{\rm pr}$ and variance $\Sigma_{\rm pr}$, applying (\ref{eq:enkfe1}) -- or
equivalently (\ref{eq:enkfe2}) -- produces a random variable $(\theta^{(1)}_k,\eta^{(1)}_k)$ 
with mean and variance given in (\ref{eq:enkfgmean}) and (\ref{eq:enkfgprec}).  Hence
the mean and variance of the ensemble members $(\theta^{(1)}_k,\eta^{(1)}_k)$ 
matches that of the Gaussian representation of the EnKF (\ref{eq:enkfgpost}).
Even though the first and second moments of the two EnKF representations
match in distribution, the ensemble representation appears to better capture
the true, right skewed posterior (compare to Figure \ref{fig:inverse}).

\subsubsection*{A two-stage approach}
\begin{figure}[h!]
  \centerline{
   \includegraphics[width=3.3in,angle=0] {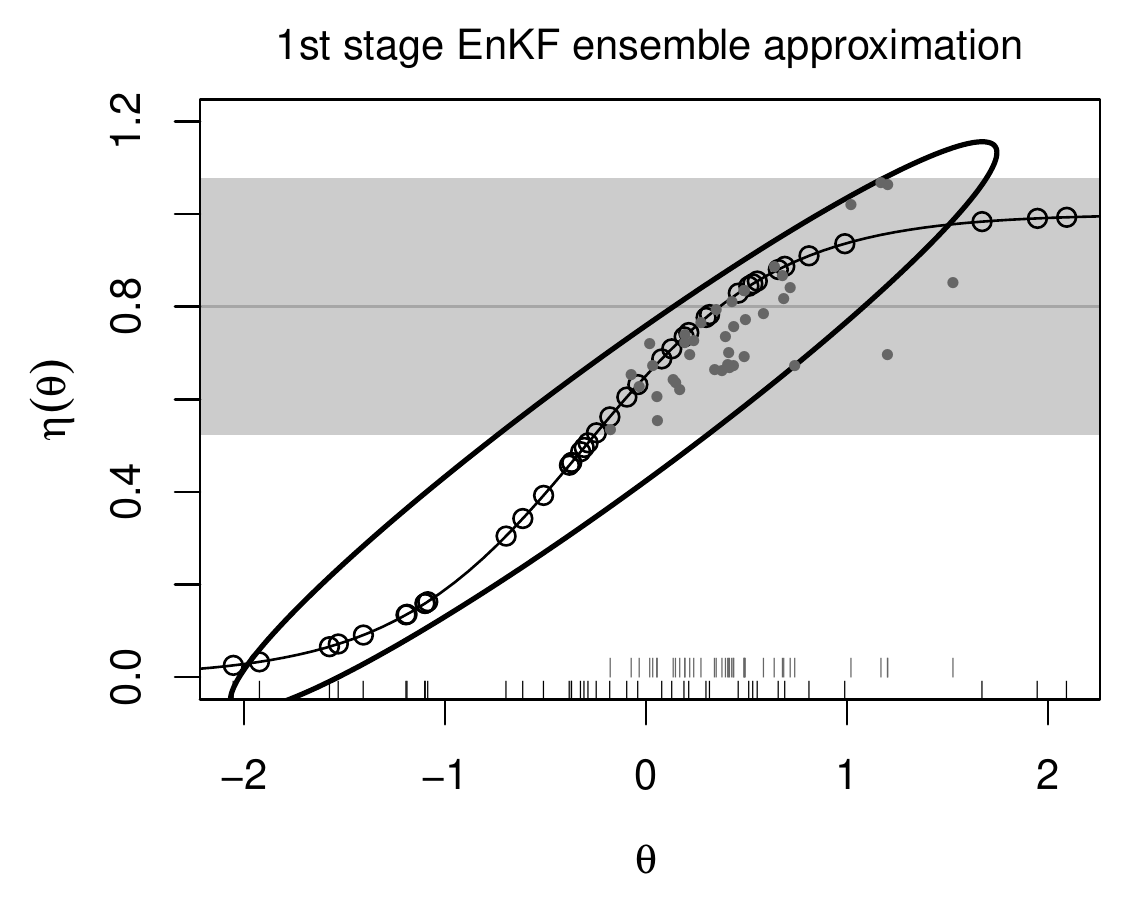}
   \includegraphics[width=3.3in,angle=0] {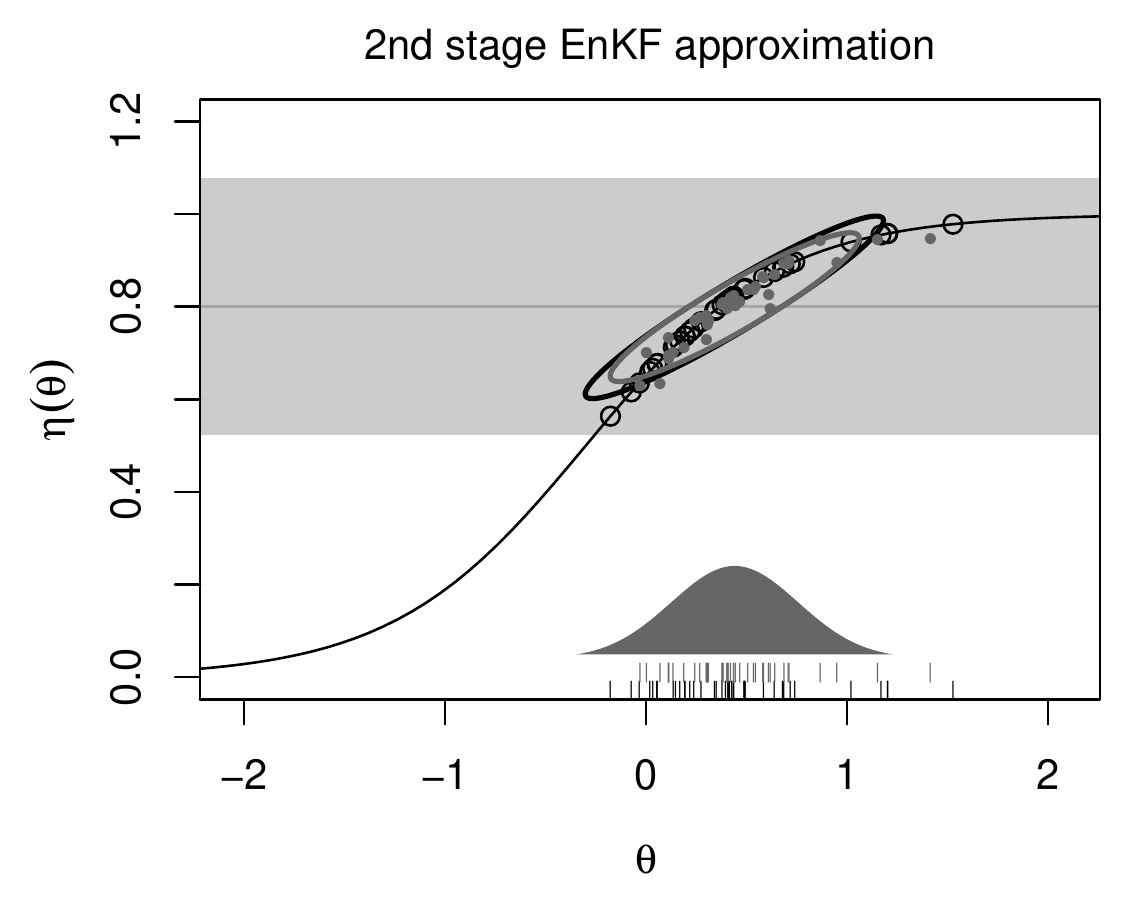}}
   \caption{\label{fig:2step} 
   A two-stage EnKF solution to the simple inverse problem.  Here the EnKF
   is applied twice, using the same observation $y$, but assuming it is observed
   with twice the variance.  Left: at the first stage, 
   the ensemble representation is used, assuming the
   observation has twice the variance, giving a data uncertainty that is 
   a factor of $\sqrt{2}$ larger than in the previous figures.  Right: the second
   stage starts with the updated
   ensemble, evaluatingng the forward model at each $\theta^{(1)}_k$, 
   producing a new ensemble 
   $(\theta^{(1)}_1,\eta(\theta^{(1)}_1)),\ldots,(\theta^{(1)}_m,\eta(\theta^{(1)}_m))$.  
   This new ensemble is updated once more, again
   using the data with twice the variance.  The resulting uncertainty
   can be represented with a Gaussian distribution (shaded density) or an ensemble
   $(\theta^{(2)}_1,\eta^{(2)}_1),\ldots,(\theta^{(2)}_m,\eta^{(2)}_m)$ (shaded dots). }
\end{figure}

Figure \ref{fig:2step} shows how one can repeatedly apply the EnKF to 
improve the accuracy of the of the normal representation of $\eta(\theta)$ where
the posterior mass for $\theta$ is concentrated.  This iterative strategy is 
closer to the original use of the EnKF for state-space estimation in non-linear,
dynamic systems. Also, this two stage approach easily generalizes to additional stages.

For this two-stage EnKF, we artificially break the information from the likelihood into
two even pieces
\[
 L(y|\eta(\theta)) \propto 
 \exp\left\{-\frac{1}{2}\frac{1}{ (2 \sigma_y^2)} (y-\eta(\theta))^2 \right\} \times
 \exp\left\{-\frac{1}{2}\frac{1}{ (2 \sigma_y^2)} (y-\eta(\theta))^2 \right\}
\]
as if $y$ were observed twice, with twice the error variance.
Then the EnKF is first applied to one of these $y$ values, with twice
the error varriance, producing an
ensemble representation $\theta^{(1)}_1,\ldots,\theta^{(1)}_m$ of the 
posterior distribution for $\theta$ given this partial piece of information.
Next, the forward model is run again at each of these new parameter settings,
producing the ensemble $(\theta^{(1)}_k,\eta(\theta^{(1)}_k)), \;k=1,\ldots,m$.
This new ensemble is now the starting point for a second EnKF update,
again using $y$ with twice the error variance.  

This second update can produce a Gaussian representation
(the gray ellipse in the right frame of Figure \ref{fig:2step}), or an 
ensemble representation  $(\theta^{(2)}_k,\eta^{(2)}_k), \;k=1,\ldots,m$
(the gray dots in the right frame of Figure \ref{fig:2step}).
As can be seen in 
the right frame of Figure \ref{fig:2step},
the second Gaussian representation of the relationship between $\theta$ and
$\eta(\theta)$ is more accurate because the $(\theta^{(1)},\eta(\theta^{(1)}))$ 
ensemble covers a narrower range, over which $\eta(\theta)$ is more nearly
linear.

Clearly, the choice of using two even splits of the likelihood information is
somewhat arbitrary -- both the number of splits and the partitioning of information
to each split could be made in many ways.  The cost of additional 
forward model evaluations has to be weighed against
the benefits of a slightly more accurate Gaussian representation of $\eta(\theta)$ over
a restricted range of values for $\theta$.

\subsubsection{Embedding the EnKF into a Bayesian formulation}
As noted in a number of references 
\citep{anderson1999monte,shumway2010time,stroud2010ensemble},
the EnKF can be embedded in a likelihood or Bayesian formulation.  For this
simple inverse problem lends itself to the Bayesian formulation below,
\begin{eqnarray*}
\mbox{sampling model: }
y|\eta(\theta) &\sim& N(\eta(\theta),\sigma_y^2) \\
\mbox{prior model: }
(\theta,\eta) &\sim& N(\mu_{\rm pr}, \Sigma_{\rm pr}),
\end{eqnarray*}
or, equivalently
\begin{eqnarray*}
y|\eta(\theta) &\sim& N(\eta(\theta),\sigma_y^2) \\
\eta|\theta &\sim& N \left( \mu_{{\rm pr}2} + \Sigma_{{\rm pr}22}^{-1} 
     \Sigma_{{\rm pr}21} (\theta - \mu_{{\rm pr}1}),
     \Sigma_{{\rm pr}22}  - 
     \Sigma_{{\rm pr}21} \Sigma_{{\rm pr}11}^{-1} \Sigma_{{\rm pr}12} 
     \right) \\
\theta &\sim& N(\mu_{{\rm pr}1},\Sigma_{{\rm pr}11}).
\end{eqnarray*}

In looking at the prior specification of $\eta|\theta$ above, it's apparent that the mean is just
the linear regression estimate of $\eta$ given $\theta$.  Hence where a GP model is used
in the formulation described in Figure \ref{fig:gp}, the EnKF implicitly uses 
a linear regression-based emulator.  While this simple form can only account
for linear effects and no interactions, the tradeoff is that this 
emulator can be estimated quickly, can handle large
ensemble sizes $m$, and can handle moderately high-dimensional input parameter,
and output spaces.

The EnKF uses the initial sample of model runs
$(\theta^\circ_1,\eta(\theta^\circ_1)),\ldots,(\theta^\circ_m,\eta(\theta^\circ_m))$ to
produce the standard plug-in estimates for $\mu_{\rm pr}$ and $\Sigma_{\rm pr}$ -- 
the sample mean and covariance.  In static inverse problems, where quick turn-around
of results isn't crucial, one could specify priors for these parameters,
producing a more fully Bayesian solution.  An obvious choice might take
vague, normal prior for $\mu_{\rm pr}$, and an inverse wishart for $\Sigma_{\rm pr}$
\citep{west:harr:1997},
if $m$ is sufficiently large relative to the dimensionality of $y$ and $\theta$.  

In cases where the dimensionality
of $\mu_{\rm pr}$ and $\Sigma_{\rm pr}$ is large (much larger than the ensemble
size $m$) covariance tapering, or some other form of localization is used to 
deal with spurious correlations produced in the standard sample covariance estimate
\citep{furrer2007estimation,evensen2009ensemble, stroud2010ensemble}.  The above specification
suggests the use of variable selection 
\citep{wasserman2000bayesian,tibshirani1996regression},
or compressed sensing \citep{baraniuk2007compressive} 
 could make a viable alternative for estimating the regression
function for $\eta$ given the ensemble draws for $\theta$, producing the updated
ensemble.
Finally, we note that a bootstrap could be a useful tool for accounting for the
uncertainty in the ensemble-based estimates for $\mu_{\rm pr}$ and $\Sigma_{\rm pr}$
since it does not require any additional model runs be carried out.

\section{Applications}
This section describes three applications in the statistical analysis of computer models 
that make use of the EnKF.
The first two are
calibration examples, one taken from cosmology, the second from climate.
The last explores how this EnKF representation can be used to for experimental
design, determining optimal spatial locations at which to take ice sheet measurements.  
The goal of these examples are to suggest possible uses of EnKF ideas,
rather than providing definitive analyses in problems involving
inference with the aid of computationally demanding computer models.

\subsection{Calibration of cosmological parameters}
Perhaps the simplest cosmological model in agreement with available
physical observations (e.g. the large scale structure of the universe,
the cosmic microwave background) is 
the $\Lambda$-cold dark matter ($\Lambda$CDM) model.
This model, controlled by a small number of parameters,
determines the composition, expansion and fluctuations of the universe.

\begin{figure}[b!]
\centerline{
\includegraphics[totalheight=2.0in,angle=0]{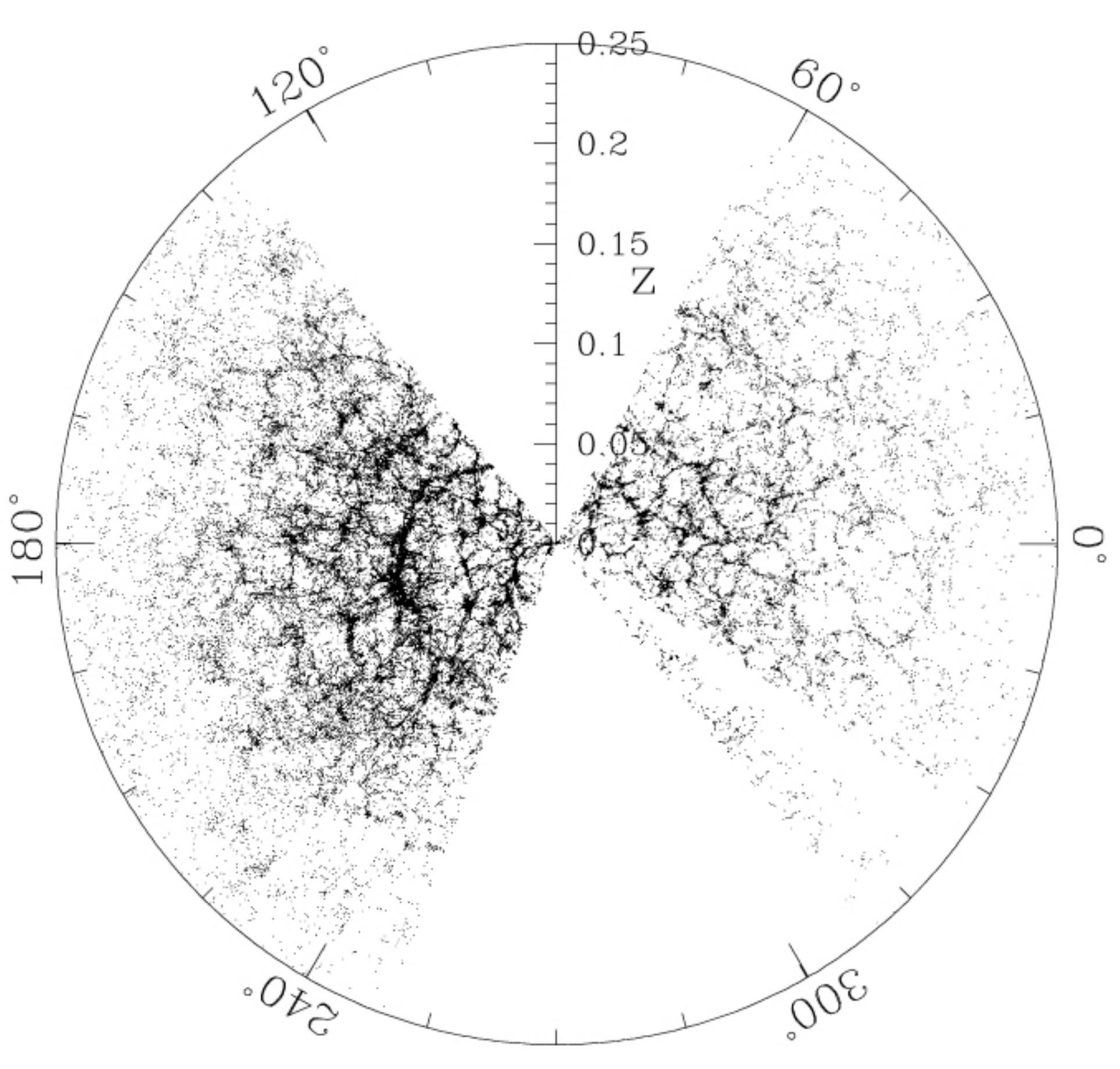}
\includegraphics[totalheight=2.0in,angle=0]{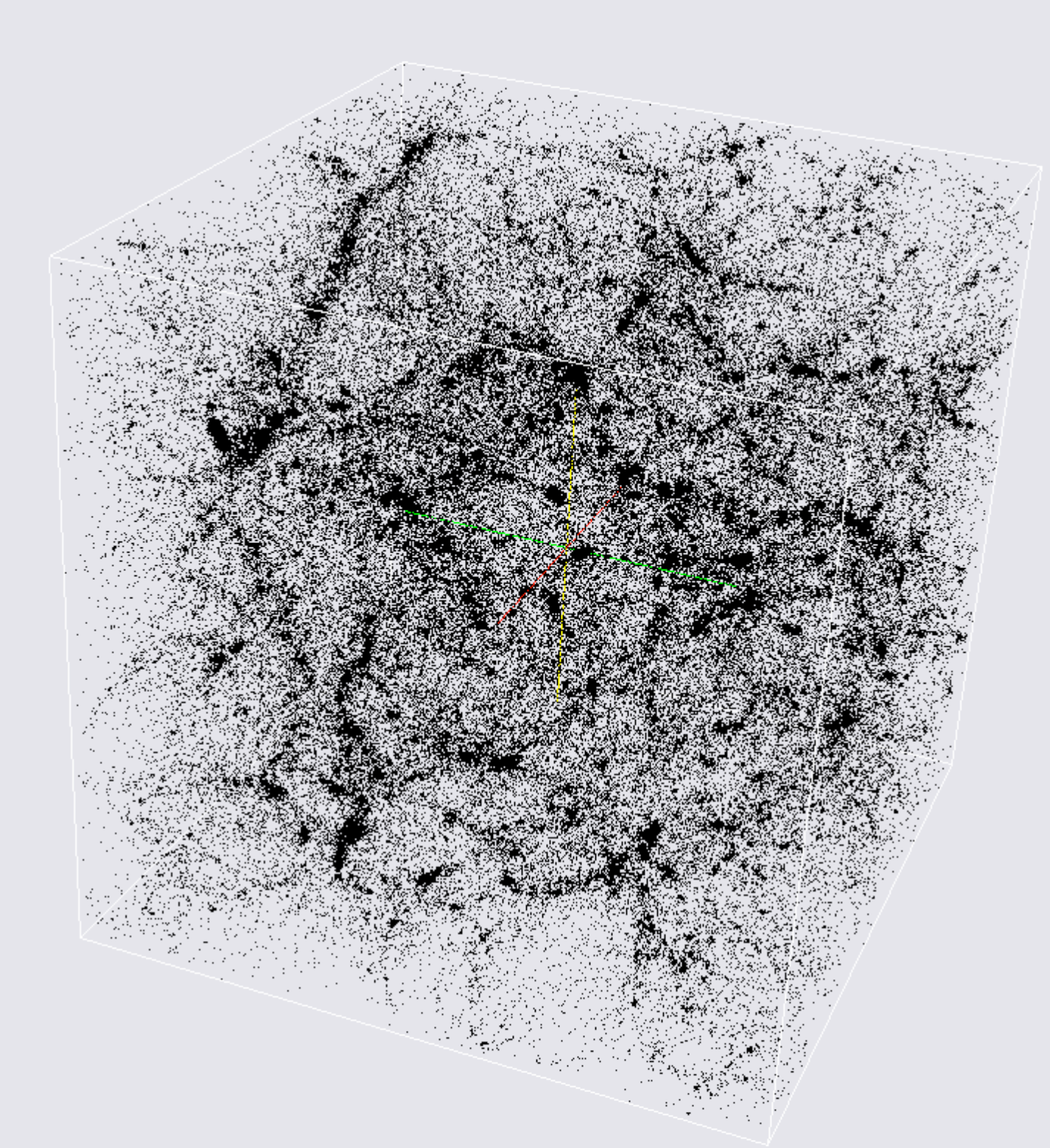}
\includegraphics[totalheight=2.0in,angle=0]{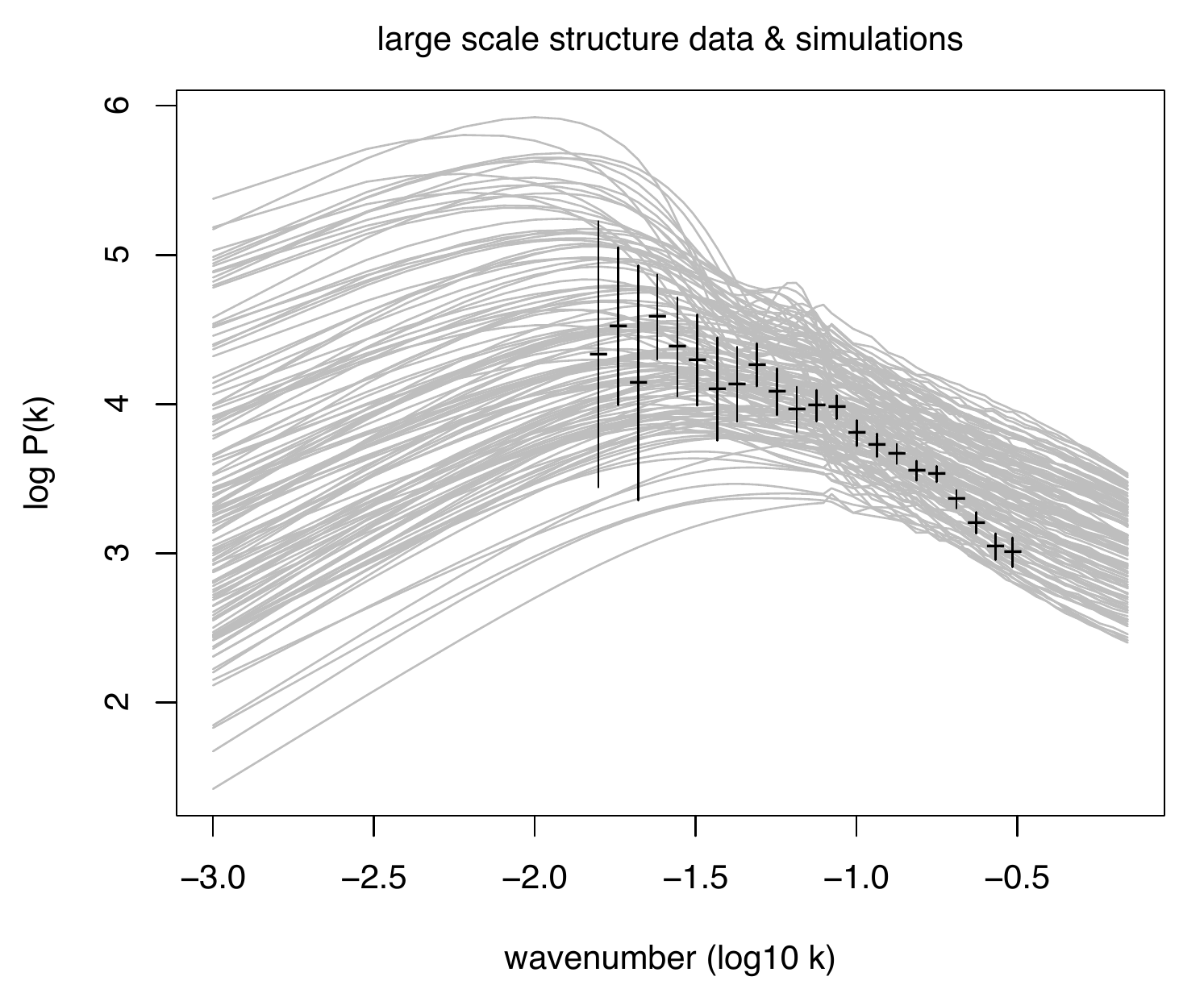}
}
\caption{\label{fig:sdss} Left: Physical observations from the Sloan Digital Sky Survey (Credit: Sloan Digital Sky Survey).  
Middle:   Simulation results from an $N$-body simulation.
Right: Power spectra for the Matter density fields.  The gray lines are from 128 simulations; the black lines give spectrum estimates derived from the physical observations.}
\end{figure}

This example focuses on model calibration, combining observations from the Sloan Digital
Sky Survey \citep{adelmanmccarthy2006fdr}, giving a local spatial map of
large galaxies, with large-scale $N$-body simulations, controlled by
five $\Lambda$CDM model parameters, evolving matter over
a history that begins with the big bang, and ends at our current time, about
14 billion years later.
An example of the physical observations produced by the SDSS are shown
in the left frame of Figure \ref{fig:sdss}.  It shows a slice of the 3-d spatial map 
of large galaxies.  Along with spatial position, the estimated mass for each of
galaxy is also recorded. 

The computational model predicts the current spatial distribution of matter in the universe, 
given the parameters of the $\Lambda$CDM model, requiring substantial 
computing effort.  For a given parameter setting, a very large-scale 
$N$-body simulation is carried out.  The simulation initializes dark matter 
tracer particles according to the cosmic microwave background and 
then propagates them according 
to gravity and other forces up to the present time.  The result of one such 
simulation is shown in the middle frame of Figure \ref{fig:sdss}.   Different 
cosmologies (i.e.~cosmological parameter settings) yield simulations 
with different spatial structure.  We would like to determine which 
cosmologies are consistent with physical observations of the SDSS
given in the left frame of Figure \ref{fig:sdss}.

Direct comparison between the simulation output and the SDSS data
is not possible since the simulations evolve an idealized, periodic cube of
particles corresponding to clusters of galaxies, while the SDSS data give a censored, local
snapshot of the large scale structure of the universe.  Instead, we
summarize the simulation output and physical observations by their
dark matter power spectra which describe the spatial distribution
of matter density at a wide range of length scales.  

Computing the matter power spectrum is trivial for the simulation
output since it is defined on a periodic, cubic lattice.  In contrast, determining
matter power spectrum from the SDSS data is a far more challenging
task since one must account for the many difficulties that accompany observational
data: nonstandard survey geometry, redshift space distortions, luminosity bias
and noise, just to name a few.  
Because of these challenges, we use the published data and likelihood of
\citet{tegmark2004tdp} which is summarized by the black lines
in the right hand frame of Figure \ref{fig:sdss}.  
The resulting data correspond to 22 pairs $(y_i,k_i)$ where $y_i$ is a binned
estimate of the log of the power, and $k_i$ denotes the wavenumber corresponding
to the estimate.  The data
vector $y=(y_1,\ldots,y_{22})'$ has a diagonal covariance 
$\Sigma_y$.  Two standard deviation error bars are shown in the right frame of
Figure \ref{fig:sdss} for each observation.

We take the ensemble produced in \cite{heit:higd:habi:2006} -- a $m=128$
run orthogonal array-based latin hypercube sample (LHS) over the 5-d 
rectangular parameter space detailed in Table \ref{tab:params}.  Since
this sample was originally generated to produce a multivariate GP emulator -- 
predicting the simulated matter power spectrum as a function of the 5-d
parameter inputs -- it is not a draw from a normal prior as is standard for 
EnKF applications.  Nevertheless, this sample can be used to estimate
$\mu_{\rm pr}$ and  $\Sigma_{\rm pr}$ from Section \ref{sec:enkf}.  The
restricted ranges of the parameters will need to be reconciled with 
the eventual normal description
of the parameter posterior, or resulting EnKF sample.

\begin{table}
\caption{$\Lambda$CDM parameters and their lower and upper bounds.}
\label{tab:params}
\centerline{
\begin{tabular}{cccc}
\hline\hline\noalign{\smallskip}
parameter & description & lower & upper \\
\noalign{\smallskip} \hline \noalign{\smallskip}
$n$ & spectral index &  0.8 & 1.4 \\
$h$ & Hubble constant & 0.5 & 1.1 \\
$\sigma_8$ & galaxy fluctuation amplitude & 0.6 & 1.6 \\
$\Omega_{\rm CDM}$ & dark matter density & 0.0 & 0.6 \\
$\Omega_{\rm B}$ & baryonic matter density & 0.02 & 0.12 \\
\noalign{\smallskip} \hline\hline
\end{tabular}
} 
\end{table}%

For each of the $m=128$ parameter settings prescribed in the LHS, the simulation
produces a 55-vector of log power spectrum outputs, given by the gray lines in the right
hand frame of Figure \ref{fig:sdss}.  Of the 55 elements in the simulation output vector, 22 of the elements are at the wavenumber $k$ corresponding to the physical observations.
Concatenating the parameter settings with the with the simulation output produces
 $m=128$ vectors of length $5+55$
\[
  \begin{pmatrix} \theta_k^\circ \\ \eta(\theta^\circ_k) \end{pmatrix},\, k=1,\ldots,m.
\]
We take $H$ to be the $22 \times 60$ incidence matrix, selecting the elements
of the vector $(\theta^\circ,\eta(\theta^\circ))$ that correspond to the physical
observations.  This, along with the physical observations $y$ and corresponding
measurement covariance $\Sigma_y$ are the necessary inputs to carry out the
Gaussian and ensemble representations of the EnKF 
described in Sections \ref{sec:enkf}.

\begin{figure}[t!]
\centerline{
\includegraphics[width=3.0in,angle=0]{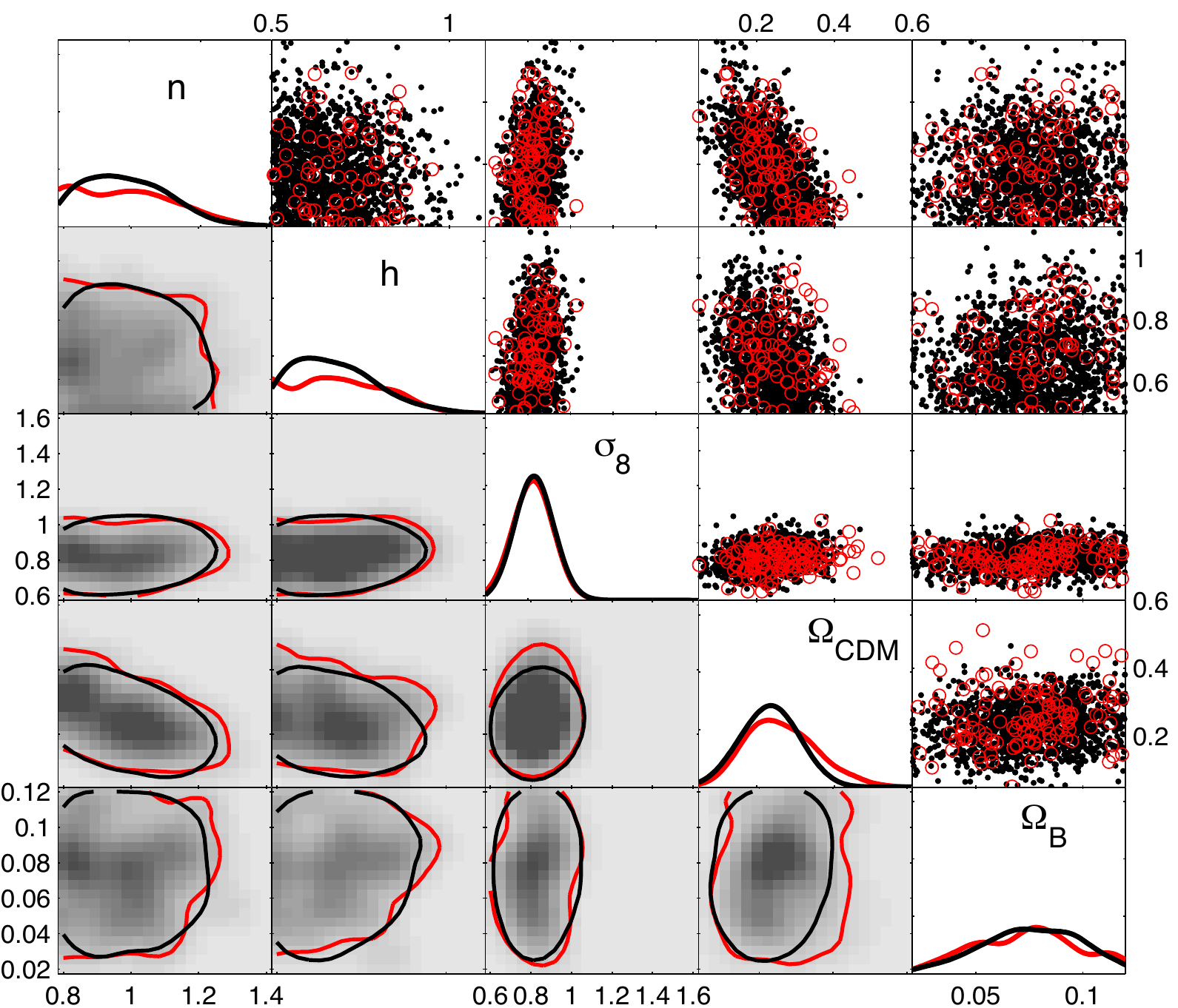}
\includegraphics[width=3.2in,angle=0]{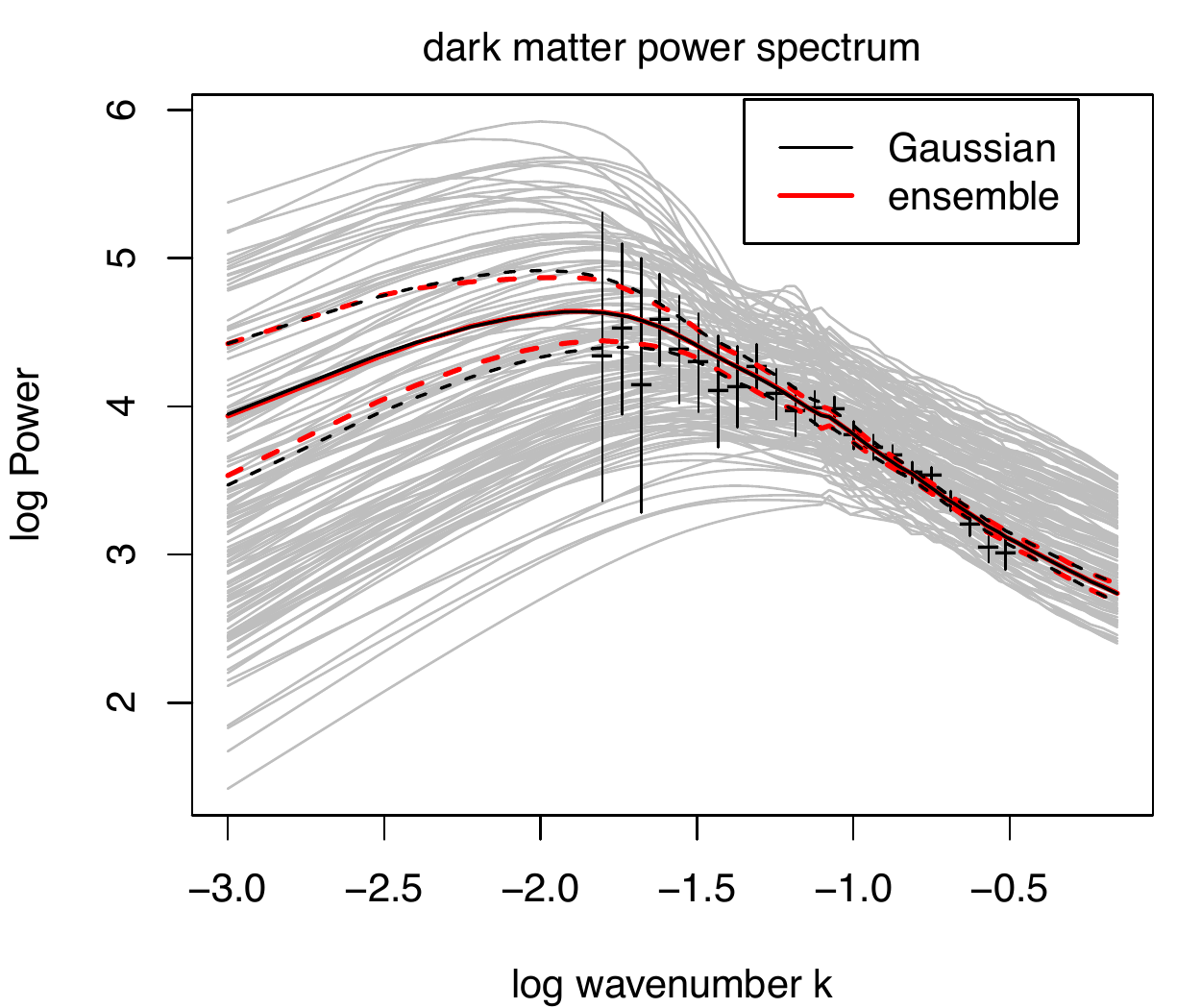}
}
\caption{\label{fig:cosmopost} The estimated posterior distribution for the model parameters
(left) and the log power (right). Left: One- and two-dimensional marginals for the estimated
posterior for the cosmological parameters.  The upper triangle shows the 128 updated
ensemble members (light circle plotting symbols) and a sample from the Gaussian representation (black dots). The lower triangle shows estimated 90\% hpd contours for
both the ensemble (light lines) and Gaussian representations (black lines).  
Right: posterior mean and pointwise 90\% credible bands for log power spectrum
for the matter density of the universe.  Light lines give the estimate produced by the
ensemble representation; black lines give the estimate produced by the 
Gaussian representation.}
\end{figure}

The estimates of the posterior distribution for the 5-dimensional parameter
vector is shown in the left frame of Figure \ref{fig:cosmopost}.  
The presence of some slight skewness
is noticible in the estimate produced by the ensemble representation.  Also produced
in these two estimation schemes is an estimate of the fitted log power spectrum, along with 
uncertainties, given in the right frame of Figure \ref{fig:cosmopost}.
These results, produced by the EnKF, can be compared to the posterior
in \citet{higd:heit:nakh:2010}, which was produced using a multivariate GP emulator,
with a far more elaborate statistical formulation.  The resulting posteriors are similar,
but both EnKF estimates seem to ``chop off'' tails in the posterior for the cosmological
parameters that are present in the GP emulator-based analysis.

\subsection{Optimal location of ice sheet measurements}
\label{sec:icesheet}
\begin{figure}[bt]
\centerline{
\includegraphics[width=5.5in,angle=0]{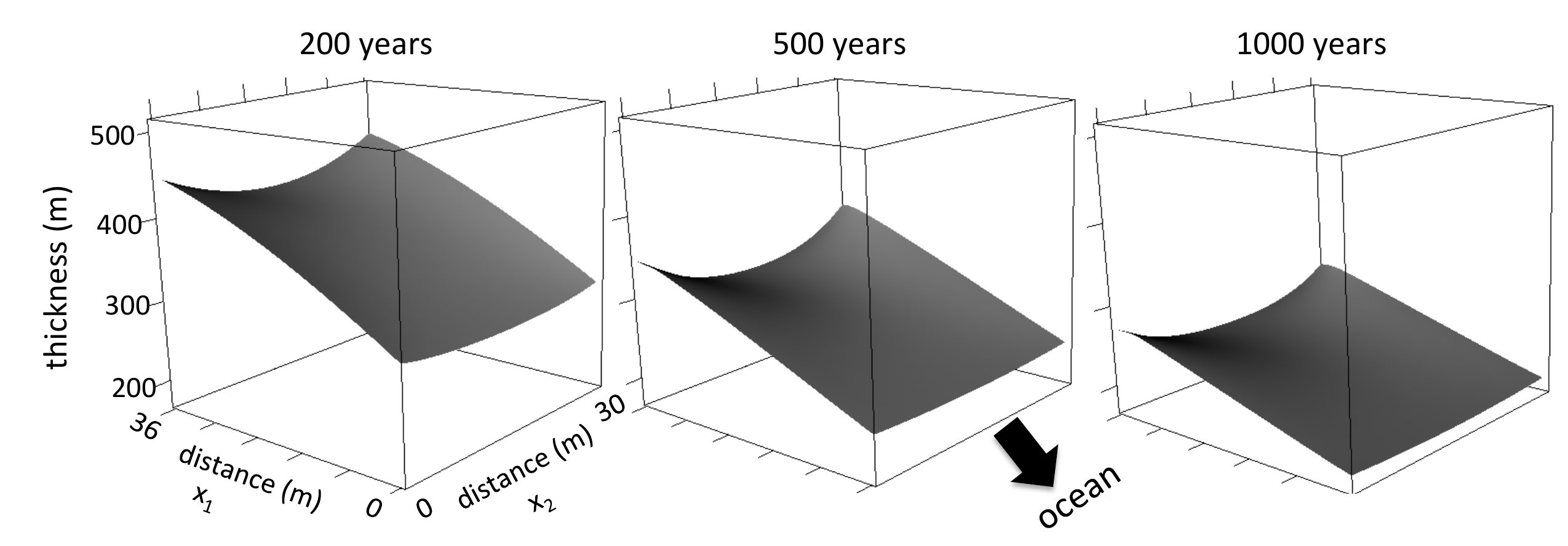}
}
\caption{\label{fig:icesheet1} Output from the idealized ice sheet model. The idealized
ice sheet is described by height over a 36m $\times$ 30m base, 
bounded on three sides by ledges.  The
fourth side is open to the ocean.  Over the span of 1000 years, the ice flows into the ocean, 
while being replenished by a prescribed precipitation.  Of interest in this application
is the thickness of the ice sheet at 1000 years.}
\end{figure}

This second application comes from an ongoing effort to use
the community ice sheet model (CISM) \citep{rutt2009glimmer,price2011committed}
along with physical measurements
to better understand ice sheet behavior and its impact
on climate.  This study considers a model of an idealized ice sheet  over a
rectangular region which is flowing out to sea on one side, while accumulating
ice from prescribed precipitation over a time of 1000 years.  This
implementation of the CISM 
depends on two parameters -- $\theta_1$ a constant in the
Glen-Nye flow law \citep{greve2009dynamics}, 
controlling the deformation of the ice sheet, 
and $\theta_2$ which controls the heat conductivity in the ice sheet.
A few time snapshots of the model output are shown in
Figure \ref{fig:icesheet1} for a particular choice of
model parameters $\theta_1$ and $\theta_2$.

While this configuration does not realistically represent important ice sheets in
Greenland or Antarctica, it is a testbed where methodology can be evaluated
for model calibration and/or planing measurement campaigns.  After 1000 years,
the thickness of the ice sheet could be measured to inform about the model
parameters $\theta_1$ and $\theta_2$.  The goal of this application is to use an ensemble of
$m=20$ model runs at different $\theta = (\theta_1,\theta_2)$ input 
settings to find a best set of 5 or 10 
locations at which to measure the ice sheet thickness.  

\begin{figure}[ht]
\centerline{
\includegraphics[width=5.0in,angle=0]{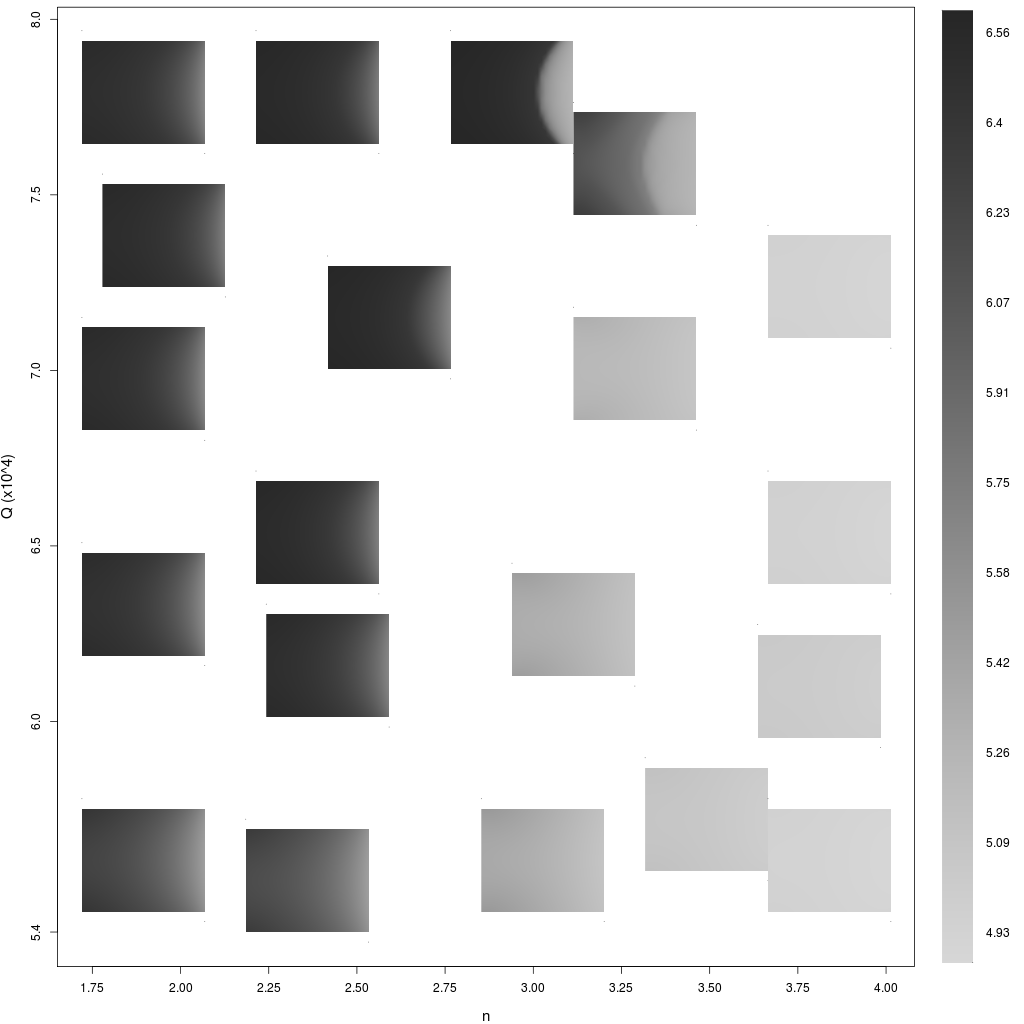}
}
\caption{\label{fig:iceEnsemble} Output from the ensemble of 20 model runs,
showing thickness of the ice sheet after 1000 years, plotted in 
the $(\theta_1,\theta_2)$-parameter
space.  $\theta_1$ controls deformation of the ice sheet; 
$\theta_2$ controls heat conductivity in the ice sheet.
Each image shows thickness as a function of spatial location; the center of the
image marks the $(\theta_1,\theta_2)$ input setting at which the model was run.
The grayscale indicates log thickness.}
\end{figure}

The parameter settings and resulting ice sheet thickness (after 1000 years) for the
$m=20$ model runs are
shown in Figure \ref{fig:iceEnsemble}.  Thickness is produced on a 
$36 \times 30$ rectangular
lattice of spatial locations.  From this figure it's clear that the
modeled ice sheet thickness is larger for smaller values of $\theta_1$, and larger
values of $\theta_2$.

The model runs produce an ensemble of $2+36\cdot 30 = p$-vectors 
$(\theta_k,\eta(\theta_k))$, $k=1,\ldots,m$.  We consider $n$ ice thickness measurements
taken at $n$ of the $36\cdot 30$ spatial grid locations given by the model.  A 
given set of $n$ measurement locations, is indexed by the $n \times p$ incidence matrix
$H$, which will contain a single 1 in each of its $n$ rows.  Thus there are
$\binom{n}{p-2}$ possible measurement designs under consideration, each 
determined by which of the last $p-2$ columns of $H$ contain a 1. 

We use the Gaussian representation of the EnKF to describe the resulting
uncertainty in $\theta$, giving a simple means to compare designs which are
determined by $H$.  Assuming the $n$ thickness measurements have 
independent measurement
errors, with a standard deviation of one meter, means $\Sigma_y$ is
the $n \times n$ identity matrix.
Then the resulting posterior variance for the joint parameter-output vector
is given by (\ref{eq:enkfgprec})
\[
\Sigma_{\rm post}^{-1} = \Sigma_{\rm pr}^{-1} + 
  H' \Sigma_y^{-1} H,
\]
with the upper $2 \times 2$ submatrix of $\Sigma_{\rm post}$ 
describing the posterior variance for the parameter vector $\theta$.

The sample covariance estimate for $\Sigma_{\rm pr}$, estimated from
only $m=20$ model runs, gives some spurious estimates for the elements
of the covariance matrix, leading to
aberrant behavior in estimates for conditional
mean and variance for $\theta$ given $\eta$.  
If we define
\begin{equation}
\label{eq:priormoments}
 \mu_{\rm pr} = 
 \begin{pmatrix} \mu_{\theta} \\
     \mu_{\eta}
\end{pmatrix}
\mbox{ and }
 \Sigma_{\rm pr} = 
 \begin{pmatrix} \Sigma_{\theta \theta} & \Sigma_{\theta \eta} \\
     \Sigma_{\eta \theta} & \Sigma_{\eta \eta}
\end{pmatrix},
\end{equation}
corresponding to the 2-vector $\theta$ and the $p-2$-vector $\eta$,
a spatial tapering covariance matrix $R(r)$ can be used
to help stabilize these estimates 
\citep{kaufman2008covariance,furrer2007estimation}
Hence we can
produce an improved estimate
\[
 \Sigma_{\eta \eta}(r) = S \circ R(r)
\]
where $S$ denotes the sample covariance matrix from the
samples $\eta_1,\ldots,\eta_m$, and
$\circ$ denotes the elementwise product of the matrix elements.
Here we take $R(r)$ to be the spatial correlation matrix induced by the isotropic exponential
correlation function, with a correlation distance of $r$.  The value for $r$ is taken
to be the maximizer of the likelihood of prior samples $\eta_1,\ldots,\eta_m$.
\[
L(r) \propto 
\prod_{k=1}^m
|\Sigma_{\eta \eta}(r)|^{-\frac{1}{2}}
\exp\left\{-\frac{1}{2}
\left( \eta_k - \mu_\eta \right)'
\Sigma_{\eta \eta}^{-1}
\left( \eta_k - \mu_\eta \right)
\right\}
\]

Using this plug-in estimate for $r$, and treating $\Sigma_{\rm pr}$ as known, 
the Bayesian $D$-optimal design
that maximizes the prior-posterior gain Shannon information is simply
the $H$ that minimizes the determinant of the $\Sigma_{\rm post}^\theta$ -- 
the upper $2 \times 2$ submatrix of $\Sigma_{\rm post}$
\citep{chaloner1995bayesian}.

Of course, since only a small number $n$ of observations are likely to be taken,
one need not compute using the full $p \times p$ matrix $\Sigma_{\rm post}$.  If
we define $H_\eta$ to be the $n \times (p-2)$ restriction of $H$, removing the 
first two columns of $H$,  then the
posterior covariance matrix for $\theta$ can be written
\[
 \Sigma^\theta_{\rm post} = \Sigma_{\theta \theta} -
  \Sigma_{\theta \eta} H'_\eta (H_\eta \Sigma_{\eta \eta}(r) H'_\eta + \Sigma_y)^{-1} 
  H_\eta \Sigma_{\eta \theta}.
\]
Here the computations require only the solve of a relatively small $n \times n$ system.
\begin{figure}[t!]
\centerline{
\includegraphics[width=5.0in,angle=0]{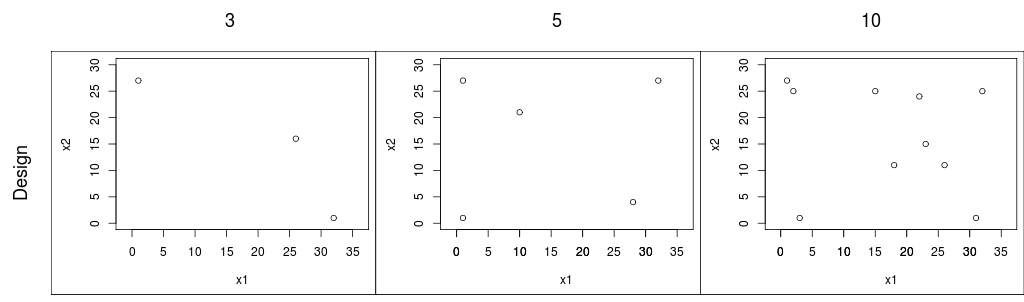}
}
\caption{\label{fig:iceDesign} Estimates of the Bayesian $D$-optimal
designs for locations at which to measure the ice sheet depth for $n=3$,
5, and 10.  The estimates use a plug-in estimate for the spatial covariance
distance of the covariance taper matrix, and Federov's exchange algorithm to
carry out the optimization.}
\end{figure}

We use the exchange algorithm of \citet{fedorov1972theory} to search for
the design $H_\eta$ that approximately minimizes the determinant 
of $\Sigma^\theta_{\rm post}$.  The estimated optimal sampling locations
for depth measurements with $n=3$, 5, and 10, are shown in Figure
\ref{fig:iceDesign}.  While the optimization algorithm only guarantees a local maximum,
we tried a large number of restarts, with the configurations giving the
minimal determinant of $\Sigma_{\rm post}^{\theta}$ shown in
Figure \ref{fig:iceDesign}.  

\subsection{Calibration of parameters in the Community Atmosphere Model}
\begin{figure}[t!]
\centerline{
\includegraphics[width=6.0in,height=6in,angle=0]{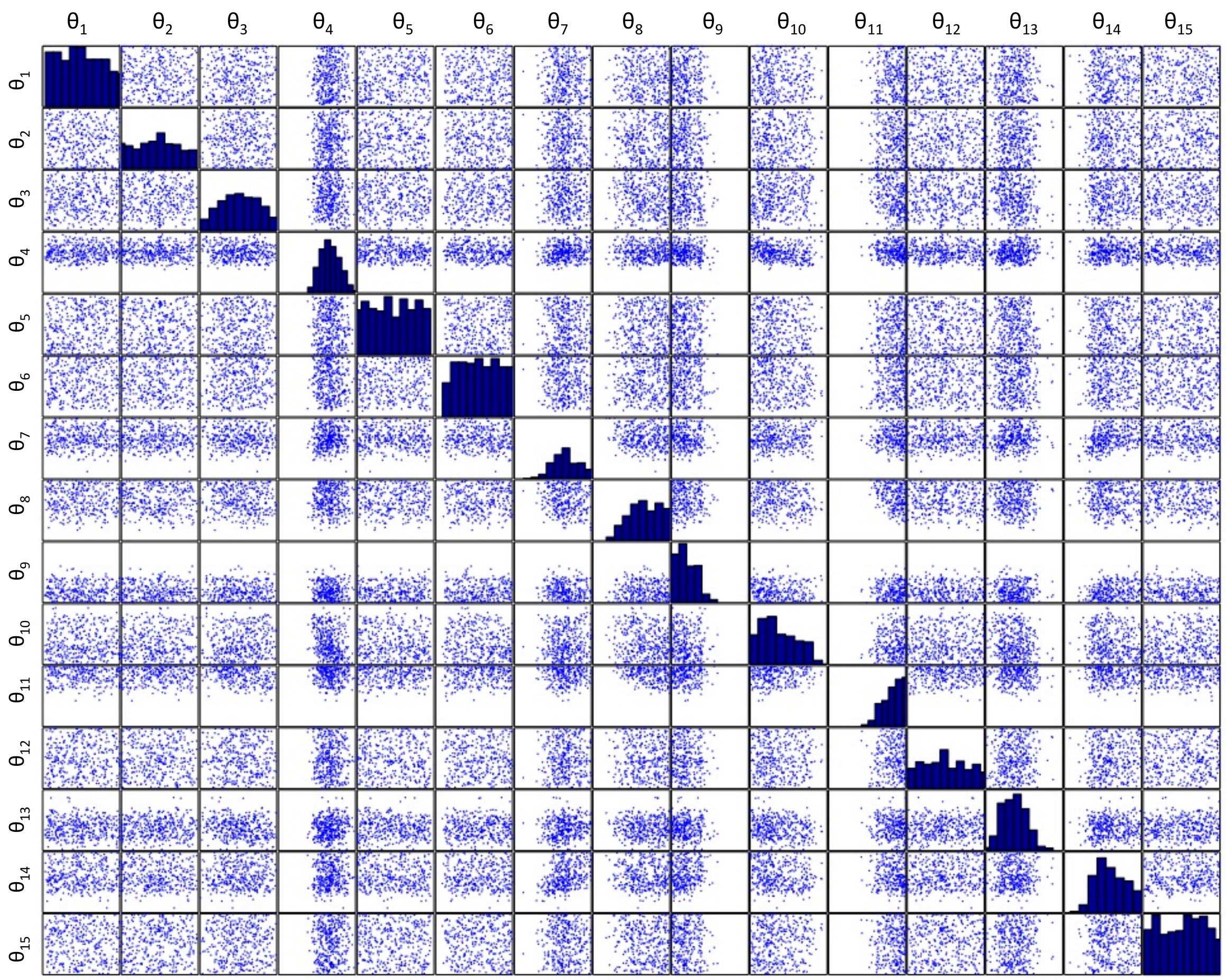}
}
\caption{\label{fig:postClimate} Ensemble representation for the posterior
distribution of the 15 model parameters after conditioning on 9 data fields.
The parameter settings for the initial ensemble are uniform over the 15-dimensional
rectangle depicted here.  
}
\end{figure}

This final application is an adaptation of the application described in
\citet{jackson2008error}, in which multiple very fast simulated annealing
(MVFSA) was used to approximate the posterior distribution of climate model parameters.
Here we use the EnKF to carry out model calibration, 
considering a more recent ensemble of 1,400 model runs, using the community
atmosphere model CAM 3.1, as described in \citet{jackson2008error}.  In this 
application 15 model parameters are sampled uniformly over a 15-dimensional rectangle
whose ranges are apparent in Figure \ref{fig:postClimate}. The
model parameters, output fields, and corresponding physical observation 
fields are listed in Table \ref{tab:climate}.
%
\begin{table}[h!t]
\caption{Climate model inputs, outputs and physical data}
\label{tab:climate}
\centerline{
\footnotesize
\begin{tabular}{clccl}
\hline \hline \noalign{\smallskip}
\multicolumn{2}{c}{inputs} && \multicolumn{2}{c}{outputs and physical data} \\
\noalign{\smallskip} \cline{1-2} \cline{4-5} \noalign{\smallskip}
 & \multicolumn{1}{c}{description} &&  &  \multicolumn{1}{c}{description} \\
 \noalign{\smallskip} \cline{2-2} \cline{5-5} \noalign{\smallskip}
$\theta_1$ & effective radius of liquid cloud droplets over sea ice &&
  $\eta_1,y_1$ & shortwave cloud forcing \\
$\theta_2$ & cloud particle number density over ocean \& land &&
  $\eta_2,y_2$ & precipitation over ocean \\
$\theta_3$ & effective radius of liquid cloud droplets over land &&
  $\eta_3,y_3$ & two meter air temperature \\
$\theta_4$ & time scale for consumption rate of deep CAPE &&
  $\eta_4,y_4$ & zonal winds at 300mb \\
$\theta_5$ & cloud particle number density over warm land &&
  $\eta_5,y_5$ & vertically averaged relative humidity \\
$\theta_6$ & threshold for autoconversion of warm ice &&
  $\eta_6,y_6$ & air temperature \\
$\theta_7$ & threshold for autoconversion of cold ice &&
  $\eta_7,y_7$ & latent heat flux over ocean \\
$\theta_8$ & effective radius of liquid cloud droplets over ocean \\
$\theta_9$ & environmental air entrainment rate \\
$\theta_{10}$ & initial cloud downdraft mass flux \\
$\theta_{11}$ & low cloud relative humidity \\
$\theta_{12}$ & ice fall velocities \\
$\theta_{13}$ & low cloud relative humidity \\
$\theta_{14}$ & deep convection precipitation efficiency \\
$\theta_{15}$ & cloud particle number density over sea ice \\
\noalign{\smallskip} \hline\hline
\end{tabular}
} 
\end{table}%

\begin{figure}[t!]
\centerline{
\includegraphics[width=6.0in,angle=0]{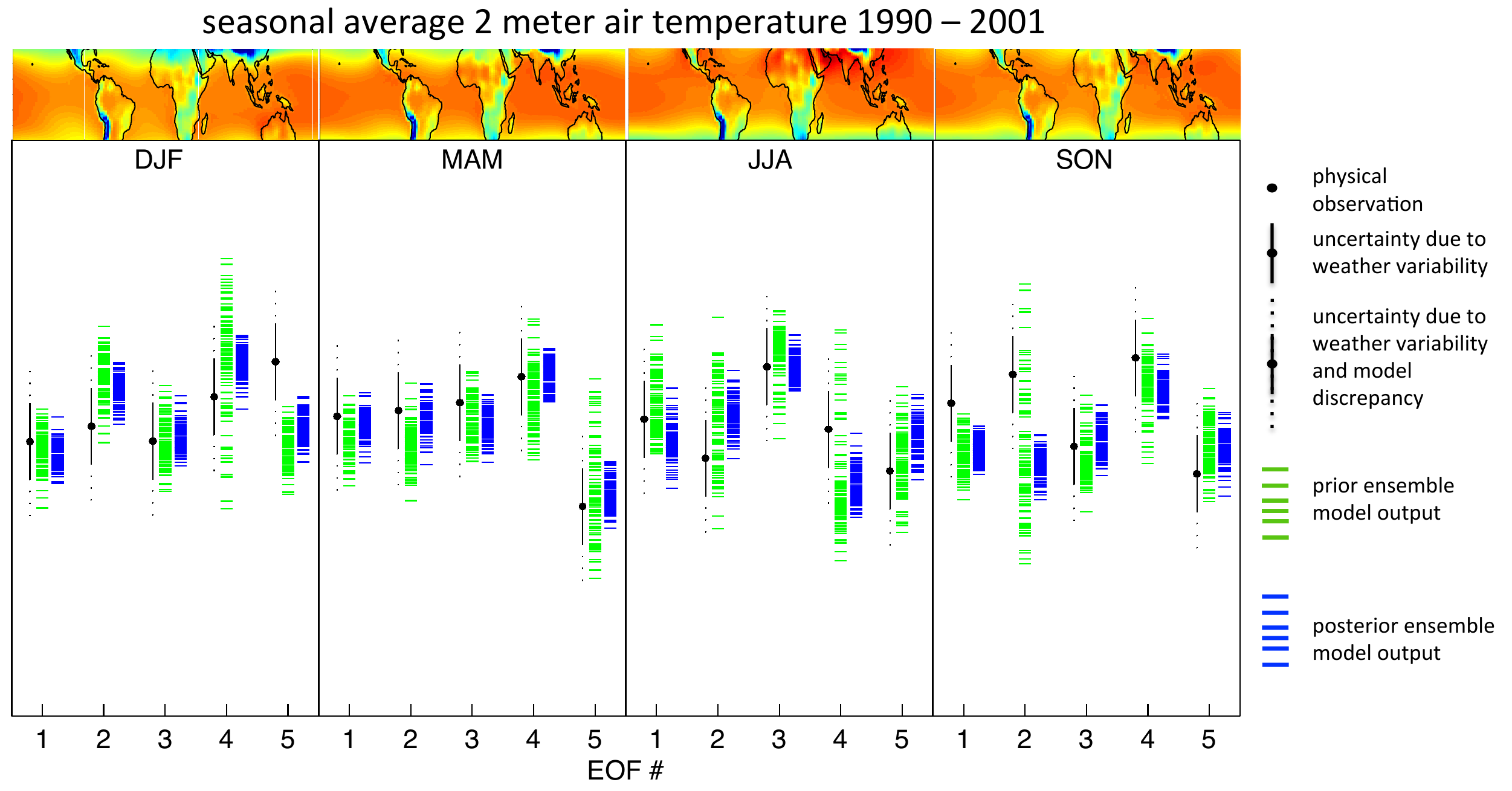}
}
\caption{\label{fig:temp2mair} Physical observations and uncertainty (black), prior simulations (light/green), and posterior
predictions (dark/blue) for seasonal 2-meter air temperature, averaged from 1990 to 2001.  The averages are computed
for each season (DJF = Dec, Jan, Feb, and so on).  The observed and simulated temperature fields are projected onto 
five EOF basis functions for each season, producing five EOF weights for each field. The basis was estimated using a single, long pilot run.  The black dot shows EOF weights corresponding to 
the physical observations, the solid black line shows a 2-$\sigma_{\rm clim}$ bound for climate variation computed from the 
pilot run; the dashed black lines show additional uncertainty due to the estimated discrepancy error.  The light/green
lines are a sample of outputs from the ensemble of model runs.  The dark/blue lines give the corresponding ensemble
representation for the updated (i.e. posterior) model predictions.  The scale of the $y$-axis has been standardized so that the 
estimated climate variance is one for each of the basis weights.  The images above the plots show the
physically observed two-meter air temperature fields.
}
\end{figure}

The computational model, described in detail in \citet{jackson2008error}, produces
a large number of outputs that could be compared to physical observations.  We focus
on a subset of the outputs (listed in Table \ref{tab:climate}) explored
in the original investigation.  Each of these outputs is recorded
as a field over the globe, averaged over 11 years (from 1990 to 2001), separately for
each season (December -- February, DJF; March -- May, MAM; June -- August, JJA; 
September -- November, SON) .  The images in Figure \ref{fig:temp2mair} show the two-meter
air temperature observations.

Rather that work directly with the model output and observed fields, we project these fields onto
a precomputed empirical orthogonal function (EOF) basis, producing a small vector of weights
-- one for each basis function -- to represent each field \citep{eof:1999}.  As in \citet{jackson2008error},
the EOF bases are computed from a long pilot run, separately from any of the model runs
used to make the ensemble.  The resulting weights for the two-meter air temperature are
shown by the light/green dashes (model) and the black dots (observation) in 
Figure \ref{fig:temp2mair}.  We use 5 EOF basis elements for each output-season
combination.  Thus the model output $\eta$ and observation fields $y$ are each 
summarized by a $7 \times 4 \times 5$ vector of weights, 
corresponding to output, season, and EOF basis respectively.

The long pilot run is also used to estimate the variation in the outputs expected just due to variation
in climate.  Thus for each output, season, and EOF, a variance $\sigma^2_{\rm clim}$ is also
estimated.  We scale the EOF bases so that each $\sigma^2_{\rm clim}$ is estimated to
be 1.  Thus, the error bars in Figure \ref{fig:temp2mair} are $\pm 2$ because of this scaling.
This scaling also makes the actual values of the $y$-axis in the figure 
essentially meaningless.

Even with this variation estimated from the pilot run, it is expected that there will still be
a discrepancy between the physical observations and the model output, even at the
best parameter setting $\theta$, for at least some of the outputs.  Hence
we specify $\Sigma_y$ to be the sum of the variance due to climate variation $I_{140}$ and a
diagonal covariance matrix that accounts for this additional discrepancy $\Sigma_\delta$.  
For each output $i$
we allow a different precision $\lambda_i$ for the discrepancy that is common across seasons and
EOF bases.  This gives
\[
\Sigma_\delta = \mbox{diag} \left( \lambda_1^{-1},\ldots,\lambda_7^{-1}\right) \otimes I_{20}
\]
so that
\[ 
 \Sigma_y = I_{140} +  \Sigma_\delta.
\]
The black dotted lines in Figure \ref{fig:temp2mair} show this additional uncertainty
due to model discrepancy for the 2 meter air temperature, governed by $\lambda_3$.  
We specify independent
$\Gamma(a=1,b=.001)$ priors for each $\lambda_i$, $i=1,\ldots,7$.

In order to estimate these precision parameters, we note that the full 140-dimensional observation
vector $y$ is modeled as the sum of normal terms
\[
y = \eta + \epsilon_{\rm clim} + \epsilon_{\rm discrep}
\]
where $\eta \sim N(\mu_\eta, \Sigma_{\eta \eta})$, with 
$\mu_\eta$ and $\Sigma_{\eta \eta}$ estimated from the prior ensemble
as defined in (\ref{eq:priormoments}), $\epsilon_{\rm clim} \sim N(0,I_{140})$, and
$\epsilon_{\rm discrep} \sim N(0,\Sigma_\delta)$.  If we define
\[
V(\lambda) = \Sigma_{\eta \eta} + I_{140} + \Sigma_\delta,
\]
we get the posterior distribution for the 7-vector $\lambda$
\begin{eqnarray*}
\pi(\lambda|y) &\propto& |V(\lambda)|^{-\frac{1}{2}} \exp\left\{ -\half(y - \mu_\eta)'
  V(\lambda)^{-1} (y-\mu\eta) \right\} \\
  && \times
  \prod_{i=1}^7 \lambda_i^{a-1} e^{-b \lambda_i}.
\end{eqnarray*}
We use the posterior mean as plug-in estimates for $\lambda$, determining $\Sigma_y$.

Now, given the dimension reduction from using the EOF bases estimated from
the pilot run, the estimate for $\Sigma_y$, and the 1400 member ensemble of 
$15+140$-vectors $(\theta,\eta(\theta))$, and the ensemble-based estimates
$\mu_{\rm pr}$ and $\Sigma_{\rm pr}$, the updated posterior distribution for
$\eta$ and $\theta$ is computed using the ensemble representation.  
The dark/blue dashes in Figure \ref{fig:temp2mair} show the posterior
ensemble for the model outputs in the EOF weight space for the two meter
air temperature.  Figure \ref{fig:postClimate} shows the posterior ensemble of
parameter values $\theta$.  The prior ensemble was sampled uniformly over
the 15-dimensional rectangle depicted in the figure. 

While the formulation presented here is very similar to that of \citet{jackson2008error}, 
we used an additive discrepancy covariance matrix, with different precisions for each
output type; theirs used a $\Sigma_y$ that is proportional to the estimated climate
variation.  Also, this analysis used fewer types of physical observations.  The resulting
posterior distribution for $\theta$ is similar, with a bit more posterior
spread in this analysis.  Also, the EnKF analysis requires only an ensemble of
model runs,  with no need for the sequential sampling required for MVFSA.

Finally, we point out that \citet{annan2005parameter} also use the EnKF to carry out
parameter estimation on a climate model.  That example uses a multi-stage
estimation approach, collecting observations over ten successive years.
That paper also uses synthetic observations
so that $\Sigma_y$ can be specified without the need for
estimation.

\section{Discussion}

This paper highlights a number of features of the EnKF from the perspective
of model calibration and shows examples of how it can be used in a variety
of applications.  Implicitly, the EnKF uses a multiple linear regression emulator
to model the mapping between model parameters and outputs.  This makes
it easy for this approach to handle large ensembles -- often a challenge for
approaches that use GP-based emulators -- as well as model outputs
that are noisy or random.  This also suggests regression-based approaches 
for dealing with high-dimensional input and output spaces
may be helpful in EnKF applications.

While the EnKF nominally starts with an initial ensemble from the prior distribution
for $\theta$, it's clear this prior will have little impact on the final results if the physical
observations are fairly constraining, as in the examples presented here.  The uniform
designs used in the applications here have little impact on the posterior results.

The results depend far more on specifications for covariance matrices, and how a large
covariance matrix is estimated from a relatively small ensemble of model runs.
We used likelihood and Bayesian approaches for estimation of covariance
parameters; a variety of alternative approaches exist in the literature
\citep{tippett2003ensemble,stroud2007sequential,evensen2009data,stroud2010ensemble,
kaufman2008covariance}.

The resulting posterior distribution for $\theta$ tends to chop off tails that would be
present in a more exact formulation.  This is clear from comparing the analyses of the
simple inverse problem laid out 
in Section \ref{sec:simpleInverse}.  This is largely due to the linearity of the regression based
emulator implicitly used in the EnKF.  We have also seen this phenomena
in the ice sheet and cosmology applications when comparing to calibration analyses based on
more exacting GP emulators.

Finally we note that the ability of the EnKF to quickly provide ``rough and ready'' results makes it ideal for
more computationally demanding tasks such experimental design or other optimization
problems that require many iterations of the estimation process.  The  
ice sheet application of Section \ref{sec:icesheet} is one such example.
 

\renewcommand{\baselinestretch}{1.0}

\end{document}